\documentclass[%
 reprint, superscriptaddress,
 amsmath,amssymb, aps, prb,
]{revtex4-1}
\usepackage{booktabs}
\usepackage{color}
\usepackage{graphicx}
\usepackage{dcolumn}
\usepackage{bm}
\usepackage{dcolumn}
\usepackage{gensymb}
\usepackage{siunitx}
\usepackage{etoolbox}
\usepackage{booktabs}
\usepackage{multirow}
\usepackage[table,xcdraw]{xcolor}
\usepackage{tabularx}
\usepackage{makecell}
\usepackage{lipsum}
\usepackage{amsmath}
\usepackage{footnote}
\makesavenoteenv{table}
\newcommand{\RNum}[1]{\uppercase\expandafter{\romannumeral #1\relax}}

\begin{document}

\preprint{APS/123-QED}
\title{
Machine learning prediction errors better than DFT accuracy
}


\author{Felix A. Faber} 
\affiliation{Institute of Physical Chemistry and National Center for Computational
Design and Discovery of Novel Materials, Department of Chemistry, University of Basel, Klingelbergstrasse 80, CH-4056 Basel, Switzerland}
\affiliation{Authors contributed equally}
\author{Luke Hutchison} 
\affiliation{Google, 1600 Amphitheatre Parkway,
Mountain View, CA, US - 94043 CA}
\affiliation{Authors contributed equally}
\author{Bing Huang}
\affiliation{Institute of Physical Chemistry and National Center for Computational
Design and Discovery of Novel Materials, Department of Chemistry, University of Basel, Klingelbergstrasse 80, CH-4056 Basel, Switzerland}
\author{Justin Gilmer}
\affiliation{Google, 1600 Amphitheatre Parkway,
Mountain View, CA, US - 94043 CA}
\author{Samuel S. Schoenholz}
\affiliation{Google, 1600 Amphitheatre Parkway,
Mountain View, CA, US - 94043 CA}
\author{George E. Dahl}
\affiliation{Google, 1600 Amphitheatre Parkway,
Mountain View, CA, US - 94043 CA}
\author{Oriol Vinyals}
\affiliation{Google, 5 New Street Square,
London EC4A 3TW, UK}
\author{Steven Kearnes}
\author{Patrick F. Riley}
\affiliation{Google, 1600 Amphitheatre Parkway,
Mountain View, CA, US - 94043 CA}
\author{O. Anatole von Lilienfeld}%
\email{anatole.vonlilienfeld@unibas.ch}
\affiliation{Institute of Physical Chemistry and National Center for Computational
Design and Discovery of Novel Materials, Department of Chemistry, University of Basel, Klingelbergstrasse 80, CH-4056 Basel, Switzerland}

\date{\today}

\begin{abstract}
We investigate the impact of choosing regressors and molecular representations for the construction of fast machine learning (ML) 
models of thirteen electronic ground-state properties of organic molecules. 
The performance of each regressor/representation/property combination is assessed using learning curves which report out-of-sample errors as a function of training set size
with up to $\sim$117k distinct molecules.  
Molecular structures and properties at hybrid density functional theory (DFT) level of theory used for training and testing come from the QM9 database [Ramakrishnan et al, {\em Scientific Data} {\bf 1} 140022 (2014)] and include dipole moment, polarizability, HOMO/LUMO energies and gap, electronic spatial extent, zero point vibrational energy, enthalpies and free energies of atomization, heat capacity and the highest fundamental vibrational frequency.
Various representations from the literature have been studied (Coulomb matrix, bag of bonds, BAML and ECFP4, molecular graphs (MG)), as well as newly developed distribution based variants including histograms of distances (HD), and angles (HDA/MARAD), and dihedrals (HDAD).
Regressors include linear models (Bayesian ridge regression (BR) and linear regression with elastic net regularization (EN)), random forest (RF), kernel ridge regression (KRR) and two types of  neural networks, graph convolutions (GC) and gated graph networks (GG).
Out-of sample errors are strongly dependent on the choice of representation {\em and} regressor {\em and} molecular property. 
Electronic properties are typically best accounted for by MG and GC, while energetic properties are better described by HDAD and KRR. 
The specific combinations with the lowest out-of-sample errors in the $\sim$117k training set size limit are 
dipole moment (MG/GC), 
static polarizability (MG/GG), 
HOMO/LUMO eigenvalue and gap (MG/GC), 
electronic spatial extent (BOB/KRR), 
zero point vibrational energy (HDAD/KRR), 
(free) energies and enthalpies of atomization (HDAD/KRR), 
heat capacity at room temperature (HDAD/KRR),
and highest fundamental vibrational frequency (BAML/RF). 
We present numerical evidence that ML model predictions deviate from DFT less than DFT deviates from experiment
for all properties. 
Furthermore, our out-of-sample prediction errors with respect to hybrid DFT reference are on par with, or close to, chemical accuracy.
Our findings suggest that ML models could be more accurate than hybrid DFT if 
explicitly electron correlated quantum (or experimental) data was available.
\end{abstract}

\maketitle

\section{Introduction}
Due to its favorable trade-off between accuracy and computational cost, Density Functional Theory (DFT) \cite{HK,KS} is 
the work-horse of quantum chemistry \cite{BurkePerspectives_2012jcp}---despite its well known shortcomings regarding spin-states, 
van der Waals interactions, and chemical reactions~\cite{ChemistsGuidetoDFT,CohenYang2012}. 
Failures to predict reaction profiles are particularly worrisome~\cite{SingletonDFTflawed2015}, and recent analysis casts even more doubts on the usefulness of DFT functionals obtained through parameter fitting~\cite{PerdewDFTScience2017}. 
The prospect of universal and computationally much more efficient machine learning (ML) models, trained on data from experiments or generated at higher levels of electronic structure theory such as post-Hartree Fock or quantum Monte Carlo 
(e.g.~exemplified in Ref.~\cite{delta_learning}), therefore represents an appealing alternative to DFT.
Not surprisingly, a lot of recent effort has been devoted to developing ever more accurate ML models of properties of molecular and condensed phase systems.

Several ML studies have already been published using a data set called QM9~\cite{gdb9},
consisting of molecular quantum properties for the $\sim$134k smallest organic molecules 
containing up to 9 heavy atoms (C, O, N, or F; not counting H) in the GDB-17 universe~\cite{gdb17}. 
Some of these studies have developed or used representations we  consider in this work, such as BAML (Bonds, 
angles, machine learning)~\cite{BAML}, bag of bonds (BoB)~\cite{bob,sk} and the Coulomb matrix (CM)~\cite{ML0,sk}. 
Atomic variants of the CM have also been proposed and tested on QM9~\cite{barker2016localized}.  
Other representations have also been benchmarked on QM9 (or QM7 which is a smaller but similar data set), such as 
Fourier series of radial distance distributions~\cite{OAvL_FRD}, motifs~\cite{Motif-Descriptor}, the smooth overlap of atomic positions (SOAP)~\cite{SOAP_original} in combination with regularized entropy match~\cite{SOAP_apl}, constant size descriptors based on connectivity and encoded distance distributions~\cite{const_size_desc}. 
\citet{delta_learning} introduced a $\Delta$-ML approach, where the difference between properties calculated at coarse/accurate quantum level of theories is being modeled. 
Furthermore, neural network models, as well as deep tensor neural networks, have recently been proposed and tested on the same or similar data sets~\cite{ANI_IsayevRoitberg2017,DeepTensorNN_2017}.
\citet{SEMIEMP_ML1} use such data to machine learn optimal molecule specific parameters for the OM2~\cite{OM2} semi-empirical method,
 and orthogonalization tests are benchmarked in Ref.~\cite{SEMIEMP_ML2}.

However, limited work has yet been done in systematically assessing {\em various} methods {\em and} properties on large sets of the exact same chemicals~\cite{AssessmentMLJCTC2013}.
In order to unequivocally establish if ML has the potential to replace DFT for the screening of properties, 
one has to demonstrate that ML test errors
are systematically lower than estimated DFT accuracies for all the properties available.
This study accomplishes that through a large scale assessment of unprecedented scale: 
(i) In order to approximate large training set sizes $N$, we included  13 quantum properties from up to $\sim$117k molecules in training (90\% of QM9).
(ii) We tested multiple regressors (Bayesian ridge regression (BR), linear regression with elastic net regularization (EN), random forest (RF), kernel ridge regression (KRR), neural network (NN) models graph convolutions (GC)~\cite{kearnes2016molecular} and gated graphs (GG)~{\cite{yujia}) and 
(iii) multiple representations including BAML, BOB, CM, extended connectivitiy fingerprints (ECFP4), histograms of distance, angle, and dihedral (HDAD), molecular alchemical radial angular distribution (MARAD), and
molecular graphs (MG). 
(iv) We investigated {\em all} combinations of regressors and representations, except for MG/NN which was exclusively used together because GC and GG depend fundamentally on the input representation being a graph instead of a flat feature vector.

The best models for the various properties are:
dipole moment (MG/GC), 
static polarizability (MG/GG), 
HOMO/LUMO eigenvalue and gap (MG/GC), 
electronic spatial extent (BOB/KRR), 
zero point vibrational energy (HDAD/KRR), 
atomization energy at 0 Kelvin (HDAD/KRR), 
atomization energy at room temperature (HDAD/KRR), 
enthalpy of atomization at room temperature (HDAD/KRR), 
atomization of free energy at room temperature (HDAD/KRR), 
heat capacity at room temperature (HDAD/KRR),
and the highest fundamental vibrational frequency (BAML/RF).
For training set size of $\sim$117k (90$\%$ of data set) we have found the additional out-of-sample error added by machine learning to be lower or as good as DFT errors 
relative to experiment for all properties, 
and that chemical accuracy (See table \ref{tab:order}) is reached, or in sight.

This paper is organized as follows: First we will briefly describe the methods, including data set, model validation protocols, representations, and regressors. In section III, we present the results and discuss them, and section IV concludes the paper. 

\section{Method}
\subsection{Data set}

We have used the QM9 data set consisting of $\sim$134k drug-like organic molecules~\cite{gdb9}.
Molecules in the data set consist of H, C, O, N and F, and contain up to 9 heavy atoms. 
For each molecule several properties, calculated at DFT level of theory (B3LYP/6-31G(2df,p)), were included. 
We used: Norm of dipole moment $\mu$ (Debye), norm of static polarizability $\alpha$ (Bohr$^3$), HOMO eigenvalue $\varepsilon_{\mathrm{HOMO}}$ (eV), LUMO eigenvalue $\varepsilon_{\mathrm{LUMO}}$ (eV), HOMO-LUMO gap $\Delta\varepsilon$ (eV), electronic spatial extent $\langle \mathrm{R}^2\rangle$ (Bohr$^2$), zero point vibrational energy ZPVE (eV), atomization energy at 0 Kelvin $\mathit{U}_{\rm 0}$ (eV), atomization energy at room temperature $\mathit{U}$ (eV), enthalpy of atomization at room temperature $\mathit{H}$ (eV), atomization of free energy at room temperature $\mathit{G}$ (eV), heat capacity at room temperature $\mathit{C}_{\mathrm{v}}$ (cal/mol/K), and the highest fundamental vibrational frequency $\omega_1$  (cm$^{-1}$). 
For energies of atomization ($\mathit{U}_{\rm 0}$, $\mathit{U}$, $H$ and $G$) all models yield very similar errors:
We will only discuss for $\mathit{U}_{\rm 0}$ for the remainder.
The 3053 molecules specified in Ref.~\cite{gdb9} which failed SMILES consistency tests were excluded from our study, 
as well as two linear molecules, leaving $\sim$131k molecules. 

\subsection{Model validation}

Starting from the $\sim$131k molecules in QM9 after removing the $\sim$3k molecules (see above) we have created a number of train-validation-test splits. 
We have splitted the data set into test and non-test sets and varied the percentage of data in test set to explore the effect of amount of data in error rates.
Inside the non-test set, we have performed 10 fold cross validation for hyperparameter optimization. 
That is, for each model 90\% (the training set) of the non-test set is used for training and 10\% (the validation set) is used for hyperparameter selection.
For each test/non-test split, we have trained 10 models on different subsets of the non-test set, and we report the mean error on the test set across those 10 models. 
Note that the non-test set will be referred to as training set in the results section in order to simplify discussion.

In terms of CPU investments necessary for training the respective models we note that 
EN/BR, RF/KRR, and GC/GG required minutes, hours, and multiple days, respectively. 
Using GPUs could dramatically reduce such timings. 

\subsection{DFT errors}
\label{sec:DFTERR}
To place the quality of our prediction errors in the right context, experimental accuracy estimates of DFT become desirable.
Here, we summarize literature results comparing DFT to experiments for the various properties we study. 
Where data is available, the corresponding deviation of DFT from experiment is listed in Table~\ref{tab:order}, 
alongside our ML prediction errors ({\em vide infra}).
More specifically, rough hybrid DFT error estimates for dipole moment and polarizability have been obtained from Refs.~\cite{ChemistsGuidetoDFT,hickey2014benchmarking}. 

Frontier molecular orbital energies (HOMO, LUMO and HOMO-LUMO gap) can not be measured directly. 
However, for the exact (yet unknown) exchange-correlation potential, HOMO and LUMO eigenvalues correspond
to the negative of the vertical ionization potential (IP) and electron affinity (EA), respectively~\cite{MeaningOfKSorbitals}.
In order to quantify the accuracy of DFT (B3LYP/6-31G(2df,p)) based frontier eigenvalues, we have carried out additional corresponding
calculations for a set of 28 organic molecules with up to 9 heavy atoms (not counting hydrogen) for which experimental
(adiabatic) IPs and EAs were reported for NIST~\cite{nist}, or in Ref.~\cite{parryang} (see also Table S2 in supplementary materials). 
Ignoring the (possibly severe) difference between vertical and adiabatic values, 
we thus approximate DFT errors of HOMO/LUMO energies by their respective deviation from experimentally measured IPs and EAs.
Correspondlingly, we estimate the HOMO-LUMO gap error by comparison to the experimental IP/EA gap. 

DFT RMSE estimates of ZPVE and $\omega_1$ (the highest fundamental vibrational frequency) were published in Ref.~\cite{AngelaVibrationDFTerror2004}
for a set of 41 organic molecules, with up to 6 heavy atoms (not counting hydrogen) and calculated using B3LYP/cc-pVTZ 
(assuming a normal distribution, we have calculated corresponding MAEs using a scaling factor of $\sim$0.8\cite{Geary1935MAE_RMSE_RATIO}).
 DFT errors estimates of energies of atomization have been taken from Ref.~\cite{JhonDFTEnthalpies1997}. 
 The errors were estimated from 69 organic molecules (22 hydrocarbons and 47 substituted hydrocarbons), with up to 6 heavy atoms (not counting hydrogens), and calculated using B3LYP/6-311+G(3df,2p).
Deviation of DFT (at the B3LYP/6-311g** level of theory) from experimental heat capacities were reported by \citet{DeLosCv2007}
who obtained errors of 16 organic molecules, with up to 8 heavy atoms (not counting hydrogens).
No satisfactory experimental data for electronic spacial extent has been found in the literature. 

Note, however, that one should be cautious when referring to these errors: Strictly speaking they can not be 
compared since different basis sets, molecules, and experiments were used.
Nevertheless, we feel that the quoted errors provide meaningful guidance as to what one can expect from DFT for each property.

\subsection{Representations}

\subsubsection{CM and BoB}
\label{sec:CM}

The Coulomb matrix (CM) representation\cite{ML0} is a square atom by atom matrix, where off diagonal elements are the Coulomb nuclear repulsion terms between atom pairs. 
The diagonal elements approximate the electronic potential energy of the free atoms. 
Atom indices in the CM are sorted by the $L^1$ norm of each atom's row (or column).
The Bag of Bonds (BoB)\cite{bob} representation uses exclusively CM elements, grouping them for different atom pairs into different bags, and sorting them within each bag by their relative magnitude.

\subsubsection{BAML}
\label{sec:BAML}
The recently introduced BAML (Bonds, angles, machine learning) representation can be viewed as a many-body extension of BoB\cite{BAML}. 
All pairwise nuclear repulsions are replaced by Morse/Lennard-Jones potentials for bonded/non-bonded atoms respectively. 
Furthermore, three- and four-body interactions between covalently bonded atoms are included using angular and torsional terms, respectively. 
Parameters and functional forms are based on the universal force field (UFF)\cite{uff}. 

\subsubsection{ECFP4}
\label{sec:ECFP4}

Extended Connectivity Fingerprints~\cite{rogers2010extended} (ECFP4) are a common representation of molecules in cheminformatics based studies. 
They are particularly popular for drug discovery~\cite{besnard2012automated,lounkine2012large,huigens2013ring}. 
The basic idea, typical also for other cheminformatics descriptors~\cite{TodeschiniConsonniHandbookDescriptor}
(e.g.~the {\em signature} descriptor~\cite{SignatureFaulon2003,Visco2002})
is to represent a molecule as the set of subgraphs up to a fixed diameter (here we use ECFP4, which is a max diameter of 4 bonds). 
To produce a fixed length vector, the subgraphs can be hashed such that every subgraph sets one bit in the fixed length vector to 1. 
In this work, we use a fixed length vector of size 1024. 
Note that ECFP4 is based solely on the molecular graph specifying all covalent bonds, e.g.~as encoded by SMILES strings.

\subsubsection{MARAD}
\label{sec:ARAD}

Molecular atomic radial angular distribution (MARAD) is a radial distribution function (RDF) based representation. 
MARAD is an extension of an atomic representation ARAD, consisting of  
three RDFs for distances between atom pairs, and parallel and orthogonal projections of distances in atom triplets, respectively.
Distances between two molecules can be evaluated analytically. 
Unfortunately, most regressors evaluated in this work, such as BR, EN and RF, do not rely on inner products and distances between representations. 
We remedy this issue by projecting MARAD onto bins in order to work with all regressors (apart for GG and GC which use MG exclusively).
Further technical details regarding MARAD are specified in the Supplementary materials~\cite{supplementary}.

\subsubsection{HD, HDA, and HDAD}

BOB, BAML and MARAD rely on computing functions for given interatomic distances, and/or angles, and/or torsions, 
and then either project that value onto discrete bins, or sort the values. 
As a straightforward alternative, we also investigated
representations which account directly from pairwise distances, triple-wise angles, and quad-wise dihedral angles through
manually generated bins in histograms. The resulting representations in
increasing interatomic many-body order are called HD (Histogram of distances), HDA (Histogram of distances and angles), and HDAD (Histogram of distances, angles and dihedral angles).
For any given molecule, one iterates through each atom $a_i$, producing a set of distances, angle and dihedral angle features for $a_i$.

\emph{Distance features} were produced by measuring the distance between $a_i$ and $a_j$ (for $i \neq j$) for each element pair.
	The distance features were assigned a label incorporating the atomic symbols of $a_i$ and $a_j$ sorted alphabetically (with H last), e.g. if $a_i$ was a carbon atom and $a_j$ was a nitrogen atom, the distance feature for the atom pair would be labeled C-N. 
	These labels will be used to group all features with the same label into a histogram and allow us to only count each pair of atoms once.

\emph{Angle features} were produced by taking the principal angles formed by the two vectors spanning from each atom $a_i$ to every subset of 2 of its 3 nearest atoms, $a_j$ and $a_k$.
The angle features were labeled by the element type of $a_i$, followed by the alphabetically sorted element types (Except for hydrogens, which were listed last) of $a_j$ and $a_k$.
The example where $a_i$ is a Carbon atom, $a_j$ a Hydrogen atom, $a_k$ a Nitrogen would be assigned the label C-N-H.

\emph{Dihedral angle features} 
were produced by taking the principal angles between two planes. We take $a_i$ as the origin, and for each of the four nearest neighbors in turn, labeling the neighbor atom $a_j$, and forming a vector $V_{ij} = a_i \rightarrow a_j$. 
Then all $\binom{3}{2}$ subsets of the remaining three out of four nearest neighbors of $a_i$ are chosen, and labeled as $a_k$ and $a_l$. 
This third and fourth atom respectively form two triangular faces when paired with $V_{ij}$: $\left\langle a_k, a_i, a_j \right\rangle$ and $\left\langle a_l, a_i, a_j \right\rangle$. 
The dihedral angle between the two triangular faces was calculated.
These dihedral angle features were labeled with the atomic symbol for $a_i$, followed by the atomic symbols for $a_j$, $a_k$ and $a_l$, sorted alphabetically, with the exception that hydrogens were listed last, e.g. C-C-N-H.

The features from all molecules have been aggregated for each label type to generate a histograms for each label type. 
Fig.~\ref{fig:hist_example} exemplifies this for C-N distances, C-C-C angles, and C-C-C-O dihedrals for the entire QM9 data set. 
Certain typical molecular characteristics can be recognized upon mere inspection. 
For example, the CN histogram displays a strong and isolated peak between 1.1 and 1.5 {\AA}, corresponding to occurrences of single, double, and triple bonds. 
For distances above 2 {\AA}, peaks at typical radii of second and third covalent bonding shells around N can be recognized at 2.6 {\AA} and 3.9 {\AA}. 
Also C-C-C angles can be easily interpreted: The peak close to zero and $\pi$ Rad corresponds to geometries where three atoms are part of a linear (alkyne, or nitrile) type of motif. 
The broad and largest peak corresponds to 120 and 109 degrees, typically observed in $sp^2$ and $sp^3$ hybridized atoms. 

The morphology of each histogram has then been examined to identify apparent peaks and troughs, motivated by the idea that peaks indicate structural commonalities among molecules. 
Bin centers have been placed at each significant local minimum and maximum (Shown as vertical lines in Fig.~\ref{fig:hist_example}). 
Values at 15-25 bin centers have been chosen as a representation for each label type. 
All bin center values are provided in the Supplementary Material.
For each molecule, the collection of features has subsequently been rendered into a fixed-size representation, 
producing one vector component for each bin center, within each label type. This has been accomplished follwoing a two-step process.
(i) \emph{Binning and interpolation:} Each feature value is projected on the two nearest bins.
 The relative amount projected on each bin uses linear projection between the two bins.
 For example: A feature with value $1.7$ which lies between two bins placed at $1.0$ and $2.0$ respectively,
 contributes $0.3$ and $0.7$ to the first and second bin respectively.
(ii) \emph{Reduction:} The collection of contributions within each bin of each molecule's feature vector is condensed to a single value by summing all contributions. 


\begin{figure*}
 \includegraphics[width=0.85\linewidth]{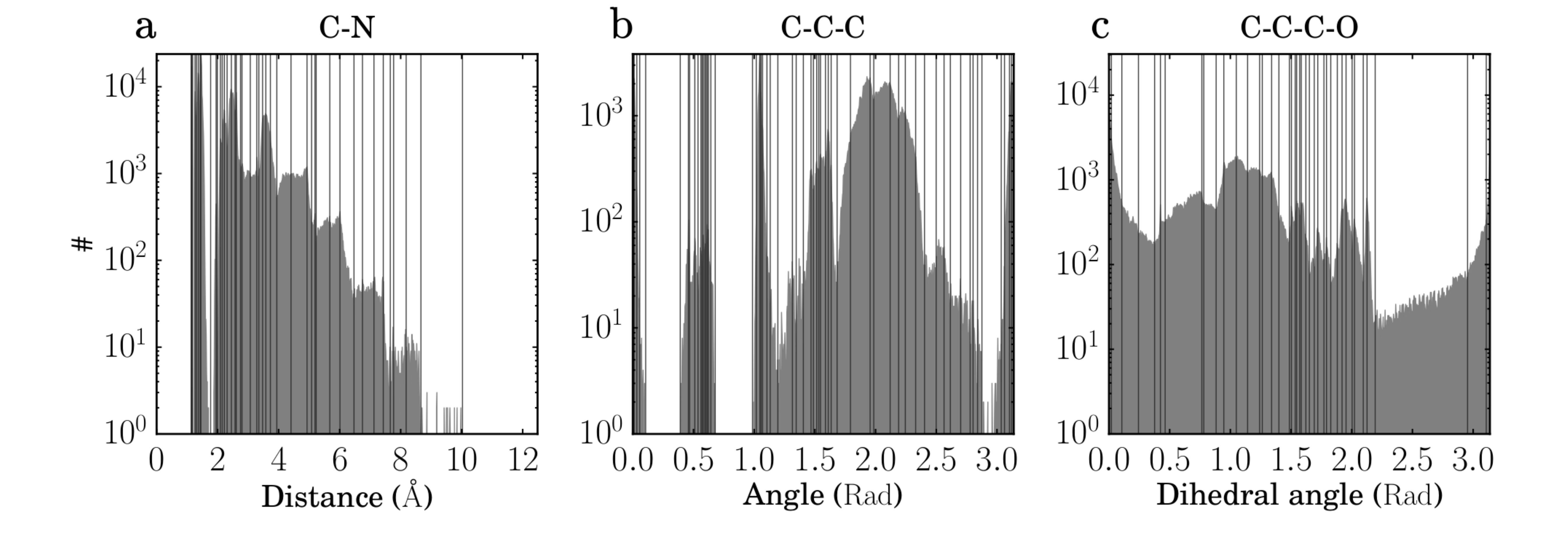}
 \caption{ \label{fig:hist_example}
 	Illustration of select histograms of distances, angles and dihedral angles in QM9. Vertical lines constitutes placements of the bins in the HD and/or HDAD representations. (a)  All C-N distances.  (b) All C-C-C angles. (c) All C-C-C-O dihedral angles.}
\end{figure*}

\subsubsection{Molecular Graphs}

We have investigated several neural network models which are based on molecular graphs (MG) as representation. 
The inputs are real-valued vectors associated with each atom and with each pair of atoms. 
More specifically, we have used the featurization described in~\citet{kearnes2016molecular} with the removal of partial charge and the addition of Euclidean distances to the pair feature vectors. 
All elements of the feature vector are described in Tables~\ref{table:atom_features} and~\ref{table:pair_features}. 

The featurization process was unsuccessful for a small number of molecules (367) because of conversion failures from geometry to rational SMILES string when using OpenBabel~\cite{obabel} or RDKit~\cite{landrum2014rdkit}, and were excluded from all results using the molecule graph features.

\begin{table}[htbp]
    \caption{Atom features for the MG representation: Values provided for each atom in the molecule.}
    \label{table:atom_features}
    \centering
    \begin{tabular}{ l| p{5cm} }
    \toprule
    Feature & Description \\\hline\hline
    \midrule
    Atom type & H, C, N, O, F (one-hot). \\\hline
    Chirality & R or S (one-hot or null). \\\hline
    Formal charge & Integer electronic charge.  \\\hline
    Ring sizes & For each ring size ($3$--$8$), the number of rings that include
                 this atom.  \\\hline
    Hybridization & $sp$, $sp^2$, or $sp^3$ (one-hot or null).  \\\hline
    Hydrogen bonding & Whether this atom is a hydrogen bond donor and/or
                       acceptor (binary values).  \\\hline
    Aromaticity & Whether this atom is part of an aromatic system. \\
    \bottomrule
    \end{tabular}
\end{table}

\begin{table}[htbp]
    \caption{Atom pair features for the MG representation: Values provided for each pair of atoms in the molecule.}
    \label{table:pair_features}
    \centering
    \begin{tabular}{ l| p{6cm} }
    \toprule
    Feature & Description  \\\hline\hline
    \midrule
    Bond type & Single, double, triple, or aromatic (one-hot or null). \\\hline
    Graph distance &For each distance ($1$--$7$), whether the shortest path between the atoms in the pair is less than or equal to that number of bonds (binary values). \\\hline
    Same ring & Whether the atoms in the pair are in the same ring.  \\\hline
    Spatial distance & The euclidean distance between the two atoms.  \\
    \bottomrule
    \end{tabular}
\end{table}

Note that within a previous draft of this study~\cite{faber2017fast}, we reported biased results for GC/GG models due to use of Mulliken partial charges within the MG representation.
All MG results presented herewithin have been obtained without any Mulliken charges in the representation. 
Model hyper parameters for the GC model, however, still correspond to the previously obtained hyper parameter search. 

\subsection{Regressors}

For all methods, we first standardized the property values so that all properties have zero mean and unit standard deviation.

\subsubsection{Kernel Ridge Regression}

KRR \cite{muller2001introduction,scholkopf2002learning,Vovk2013,kernel_ridge_regression2}  is a type of regression with regularization
\cite{Ridge_Regression} which uses kernel functions as basis set.
A property $p$ of a query molecule $\mathbf{m}$ is predicted by a sum of weighted kernel functions $K(\mathbf{m},\mathbf{m}^{\mathrm{train}}_i)$ between $\mathbf{m}$ and all molecules $\mathbf{m}^{\mathrm{train}}_i$ in the training set, 

\begin{eqnarray}
\label{eq:krr_predicting}
p(\mathbf{m}) & = & \sum_i^N  \alpha_i K(\mathbf{m},\mathbf{m}^{\mathrm{train}}_i)
\end{eqnarray}
where $\alpha_i$ are regression coefficients, obtained by minimizing the Euclidean distance between the estimated and the reference property of all molecules in the training set. 
We used Laplacian and Gaussian kernels as implemented by scikit-learn~\cite{scikit-learn} for all representations.

The level of noise in our data is very low so strong regularization is not necessary. We therefore set the regularization parameter to $10^{-9}$. Kernel widths were chosen by screening values on a base-2 logarithmic grid for the 10 percent training set (from 0.25 to 8192 for Gaussian kernel and 0.1 to 16384 for Laplacian kernel). In order to simplify the width screening, prior to learning all feature vectors were normalized (scaling the input vector by the mean norm across the training set) by the Euclidean norm for the Gaussian kernel and the Manhattan norm for the Laplacian kernel.

\subsubsection{Bayesian Ridge Regression}

We use BR~\cite{Neal:1996:BLN:525544} as is implemented in scikit-learn~\cite{scikit-learn}. 
BR is a linear model with a $L^2$ penalty on the coefficients. 
Unlike Ridge Regression where the strength of that penalty is a regularizing hyperparameter which must be set, 
in Bayesian Ridge Regression the optimal regularizer is estimated from the data.

\subsubsection{Elastic Net}

Also EN~\cite{RSSB:RSSB503} is a linear model. Unlike BR, the penalty on the weights is a mix of $L^1$ and $L^2$ terms. In addition to the regularizing hyperparameter for the weight penalty, Elastic net has an additional hyperparameter $\text{l1\_ratio}$ to control the relative strength of the $L^1$ and $L^2$ weight penalties.  
We used the scikit-learn~\cite{scikit-learn} implementation and set $\text{l1\_ratio}=0.5$. We then did a hyperparmaeter search on regularizing parameter in a base 10 logarithmic grid from $1\mathrm{e}-6$ to $1.0$.

\subsubsection{Random Forest}

We use RF~\cite{breiman2001random} as implemented in scikit-learn~\cite{scikit-learn}. RF regressors produce a value by averaging many individual decision trees fitted on randomly resampled sets of the training data. Each node in each decision tree is a threshold of one input feature.
Early experiments did not reveal strong differences in performance based on the number of trees used, once a minimal number was reached. 
We have used 120 trees for all regressions.

\subsubsection{Graph Convolutions}

We have used the GC model as described in~\citet{kearnes2016molecular}, with several structural modifications and optimized hyperparameters. 
The graph convolution model is built on the concepts of ``atom'' layers (one real vector associated with each atom) and ``pair'' layers (one real vector associated with each pair of atoms). 
The graph convolution architecture defines operations to transform atom and pair layers to new atom and pair layers. 
There are three structural changes to the model used herewithin when compared to the one described in~\citet{kearnes2016molecular}. 
We describe these briefly here with details in the Supplementary Material. 
First, we have removed the ``Pair order invariance'' property by simplifying the ($A \rightarrow P$) transformation.  
Since the model only uses the atom layer for the molecule level features, pair order invariance is not needed.
Second, we have used the Euclidean distance between atoms. 
In the ($P \rightarrow A$) transformation, we divide the value from the convolution step by a series of distance exponentials. 
If the original convolution for an atom pair $(a, b)$ with distance $d$ produces a vector $V$, we concatenate the vectors $V$, $\frac{V}{d^{1}}$, $\frac{V}{d^{2}}$, $\frac{V}{d^{3}}$, and $\frac{V}{d^{6}}$ to produce the transformed value for the pair $(a, b)$.
Third, we have followed other work on neural networks based on chemical graphs~\cite{duvenaud2015convolutional}. 
Inspired by fingerprinting like Extended Connectivity Fingerprints~\cite{rogers2010extended}, 
based on summing softmax operations to convert a real valued vector to a sparse vector and sum those sparse vectors across all the atoms, 
we have used the same operation here along with a simple sum across the atoms to produce molecule level features from the top atom layer. 
We have found that this works as well or better than the Gaussian histograms first used in GC~\cite{kearnes2016molecular}.
To optimize the network, we have searched the hyperparameter space using Gaussian Process Bandit Optimization~\cite{JMLR:v15:desautels14a} as implemented by HyperTune~\cite{hypertune}. 
The hyperparameter search has been based on the evaluation of the validation set for a single fold of the data.
Further details including parameters, and search ranges chosen for this paper are listed in the Supplementary materials~\cite{supplementary}.

\subsubsection{Gated Graph Neural Networks}
 We have used the GG Neural Networks model (GG) as described in~\citet{yujia}. 
Similar to the GC model, it is a deep neural network whose input is a set of node features $\{x_v, v \in G\}$, and an adjacency matrix $A$ with entries in a discrete set $S = \{0,1,\cdots,k\}$ to indicate different edge types. 
It has internal hidden representations for each node in the graph $h_v^t$ of dimension $d$ which it updates for $T$ steps of computation. 
Its output is invariant to all graph isomorphisms, meaning the order of the nodes presented to the model does not matter. 
 We have implemented the GG model as described in the original work with some additional preprocessing of the input. 
To include distance information we bin atomic distances into 10 bins, 
[0, 2],  [2,2.5], [2.5,3], [3,3.5], [3.5,4], [4,4.5], [4.5,5], [5,5.5], [5.5,6], and [6,$\infty$]
in \AA, respectively. 
Using these bins, the adjacency matrix has entries in an alphabet of size 14 (k=14), 
indicating bond type for covalently bonded atoms, and distance bin for all other atoms. 
We have trained the GG model on each target property individually. 
Further technical details are specified in the Supplementary materials~\cite{supplementary}.

\section{Results and discussion}
\label{sec:results}

\begin{table*}[h!]
\centering
\caption{
\label{tab:order}
MAE on out-of-sample data of all representations for all regressors and properties at $\sim$117k (90$\%$) training set size. Regressors include Bayesian ridge regression (BR), linear regression with elastic net regularization (EN), random forest (RF), kernel ridge regression (KRR) and molecular graphs based neural networks (GG/GC). 
The best combination for each property are highlighted in bold. 
Additionally, the table contains mean absolute deviation (MAD), target MAE, DFT (B3LYP) MAE relative to experiment for all properties in the QM9 data set; mean MAE of representations for each property and regressor; and
normalized (by MAD) mean MAE (NMMAE) over all properties for each regressor/representation combination.
}

\footnotetext{\label{note:sk} 
Using the same values for target accuracies as in \cite{sk}.
 Target accuracy for energies of atomization, and orbital energies were set to 1 kcal/mol, which is generally accepted as (or close to) chemical accuracy within the chemistry community. 
 Target accuracies used for $\mu$ and $\alpha$ are 0.1 D and 0.1 Bohr$^3$ respectively, which is within the error of CCSD relative to experiments\cite{hickey2014benchmarking}.
 Target accuracies used for $\omega_1$ and ZPVE are $10^{-1}$ cm$^{-1}$, which is slightly larger than CCSD(T) error for predicting frequencies
\cite{Wim2007Harmonic}.
 Target accuracies used for $\langle\mathrm{R}^2\rangle$ and $\mathit{C}_{\mathrm{v}}$ were not explained in article~\cite{sk}. 
 }

\footnotetext{\label{note:dft}  See section \ref{sec:DFTERR} for information on how the errors where obtained.}

{%
\robustify\bfseries
\sisetup{detect-weight=true,detect-inline-weight=math}
\begin{tabular}{l|l|
S[table-format=3.2]
S[table-format=3.2]
S[table-format=3.2]
S[table-format=3.2]
S[table-format=3.2]
S[table-format=3.2]
S[table-format=3.4]
S[table-format=3.2]
S[table-format=3.2]
S[table-format=3.2]|
S[table-format=3.2]}

\hline\hline
\multicolumn{1}{c}{} & \multicolumn{1}{c}{} & \multicolumn{1}{c}{$\mu$} & \multicolumn{1}{c}{$\alpha$} & \multicolumn{1}{c}{$\varepsilon_{\mathrm{HOMO}}$} & \multicolumn{1}{c}{$\varepsilon_{\mathrm{LUMO}}$} & \multicolumn{1}{c}{$\Delta\varepsilon$} & \multicolumn{1}{c}{$\langle\mathrm{R}^2\rangle$} & \multicolumn{1}{c}{$\mathrm{ZPVE}$} & \multicolumn{1}{c}{$\mathit{U}_{\rm 0}$} & \multicolumn{1}{c}{$\mathit{C}_{\mathrm{v}}$} & \multicolumn{1}{c}{$\omega_1$} & \multicolumn{1}{c}{NMMAE}\\\hline
\multicolumn{1}{c}{} & \multicolumn{1}{c}{} & \multicolumn{1}{c}{$\mathrm{Debye}$} & \multicolumn{1}{c}{$\mathrm{Bohr}^3$} & \multicolumn{1}{c}{$\mathrm{eV}$} & \multicolumn{1}{c}{$\mathrm{eV}$} & \multicolumn{1}{c}{$\mathrm{eV}$} & \multicolumn{1}{c}{$\mathrm{Bohr}^2$} & \multicolumn{1}{c}{$\mathrm{eV}$} & \multicolumn{1}{c}{$\mathrm{eV}$} & \multicolumn{1}{c}{$\mathrm{cal/molK}$} & \multicolumn{1}{c}{$\mathrm{cm}^{-1}$} &  \multicolumn{1}{c}{arb. u.} \\
\hline
  Mean& & 2.67 & 75.3 & -6.54 & 0.322 & 6.86 & 1190 & 4.06 &-76.6 & 31.6 & 3500 & \\
   MAD& & 1.17 & 6.29 & 0.439 & 1.05 & 1.07 & 203 & 0.717 & 8.19 & 3.21 & 238 &   \\

  Target$^{\text{\ref{note:sk}}}$ & & 0.10  & 0.10  & 0.043 & 0.043 & 0.043 & 1.2 & 0.0012  & 0.043 & 0.050 & 10 &  \\
  DFT$^{\text{\ref{note:dft}}}$ & & 0.10 & 0.4 & 2.0 & 2.6 
 & 1.2 
 & \multicolumn{1}{c}{-} & 0.0097 & 0.10 & 0.34 & 28 &  \\

\hline\hline


 
\multirow{8}{*}{EN}
 & CM   & 0.844 & 1.33  & 0.338 & 0.631 & 0.722 & 55.5 & 0.0265 & 0.911 & 0.906  & 131  & 0.423\\
 & BOB  & 0.763 & 1.20  & 0.283 & 0.521 & 0.614 & 55.3 & 0.0232 & 0.602 & 0.700  & 81.4 & 0.35\\
 & BAML & 0.686 & 0.793 & 0.186 & 0.275 & 0.339 & 32.6 & 0.0129 & 0.212 & 0.439  & 60.4 & 0.231\\ 
 & ECFP4& 0.737 & 3.45  & 0.224 & 0.344 & 0.383 & 118  & 0.270  & 3.68  & 1.51   & 86.6 & 0.462\\
 & HDAD & 0.563 & 0.437 & 0.139 & 0.238 & 0.278 & 6.19 & 0.00647& 0.0983 & 0.0876&  94.2& 0.183\\
 & HD   & 0.705 & 0.638 & 0.203 & 0.299 & 0.360 & 6.70 & 0.00949 & 0.192 & 0.195 & 104  & 0.236\\
 & MARAD& 0.707 & 0.698 & 0.222 & 0.305 & 0.391 & 27.4 & 0.00808 & 0.183 & 0.206 & 108  & 0.256\\\hline
 & Mean & 0.715 & 1.22  & 0.228 & 0.373 & 0.441 & 43.1 & 0.0509  & 0.840 & 0.578 & 95.1 & \\
 \hline \hline 
\multirow{7}{*}{BR}
 & CM & 0.844 & 1.33 & 0.338 & 0.632 & 0.723 & 55.5 & 0.0265 & 0.911 & 0.907 & 131 & 0.424\\
 & BOB & 0.761 & 1.14 & 0.279 & 0.521 & 0.614 & 48.0 & 0.0222 & 0.586 & 0.684 & 80.9 & 0.343\\
 & BAML & 0.685 & 0.785 & 0.183 & 0.275 & 0.339 & 30.4 & 0.0129 & 0.202 & 0.444 & 60.4  & 0.229\\
& ECFP4 & 0.737 & 3.45 & 0.224 & 0.344 & 0.383 & 118 & 0.270 & 3.69 & 1.51 & 86.7 & 0.462\\
 & HDAD & 0.565 & 0.43 & 0.14 & 0.238 & 0.278 & 5.94 & 0.00318 & 0.0614 & 0.0787 & 94.8 & 0.182\\
 & HD & 0.705 & 0.633 & 0.203 & 0.298 & 0.359 & 6.8 & 0.00693 & 0.171 & 0.19 & 104 & 0.235\\
 & MARAD & 0.647 & 0.533 & 0.18 & 0.257 & 0.315 & 26.8 & 0.00854 & 0.171 & 0.201 & 103 & 0.226\\\hline
 & Mean & 0.706 & 1.19 & 0.221 & 0.367 & 0.430 & 41.7 & 0.0500 & 0.828 & 0.574 & 94.5 & \\
 \hline \hline
\multirow{7}{*}{RF}

 & CM & 0.608 & 1.04 & 0.208 & 0.302 & 0.373 & 45.0 & 0.0199 & 0.431 & 0.777 & 13.2 & 0.239\\
 & BOB & 0.450 & 0.623 & 0.120 & 0.137 & 0.164 & 39.0 & 0.0111 & 0.202 & 0.443 & 3.55 & 0.142\\ 
 & BAML & 0.434 & 0.638 & 0.107 & 0.118 & 0.141 & 51.1 & 0.0132 & 0.200 & 0.451 & \bfseries 2.71 &  0.141 \\
& ECFP4 & 0.483 & 3.70 & 0.143 & 0.145 & 0.166 & 109 & 0.242 & 3.66 & 1.57 & 14.7 & 0.349\\
 & HDAD & 0.454 & 1.71 & 0.116 & 0.136 & 0.156 & 48.3 & 0.0525 & 1.44 & 0.895 & 3.45 & 0.198\\
 & HD & 0.457 & 1.66 & 0.126 & 0.139 & 0.150 & 46.8 & 0.0497 & 1.39 & 0.879 & 4.18 & 0.197\\
 & MARAD & 0.607 & 0.676 & 0.178 & 0.243 & 0.311 & 45.3 & 0.0102 & 0.21 & 0.311 & 19.4 & 0.199\\\hline
 & Mean & 0.499 & 1.43 & 0.142 & 0.174 & 0.209 & 54.9 & 0.0569 & 1.08 & 0.761 & 8.74 & \\
 \hline \hline 
 \multirow{7}{*}{KRR}
  & CM & 0.449 & 0.433 & 0.133 & 0.183 & 0.229 & 3.39 & 0.0048 & 0.128 & 0.118 & 33.5 & 0.136\\
 & BOB & 0.423 & 0.298 & 0.0948 & 0.122 & 0.148 & \bfseries 0.978 & 0.00364 & 0.0667 & 0.0917 & 13.2 & 0.0981\\
 & BAML & 0.460 & 0.301 & 0.0946 & 0.121 & 0.152 & 3.9 & 0.00331 & 0.0519 & 0.082 & 19.9 & 0.105\\
 & ECFP4 & 0.490 & 4.17 & 0.124 & 0.133 & 0.174 & 128 & 0.248 & 4.25 & 1.84 & 26.7 & 0.383\\
 & HDAD & 0.334 & 0.175 & 0.0662 & 0.0842 & 0.107 & 1.62 & \bfseries 0.00191 & \bfseries 0.0251 & \bfseries 0.0441 & 23.1 & 0.0768\\
 & HD & 0.364 & 0.299 & 0.0874 & 0.113 & 0.143 & 1.72 & 0.00316 & 0.0644 & 0.0844 & 21.3 & 0.0935\\
 & MARAD & 0.468 & 0.343 & 0.103 & 0.124 & 0.163 & 7.58 & 0.00301 & 0.0529 & 0.0758 & 21.3 & 0.112\\\hline 
 & Mean & 0.427 & 0.859 & 0.101 & 0.126 & 0.159 & 21.1 & 0.0383 & 0.662 & 0.333 &  22.7 &\\
 \hline \hline
GG & MG & 0.247 &  \bfseries 0.161 & 0.0567 & 0.0628 &  0.0877 & 6.30 & 0.00431 & 0.0421 & 0.0837 & 6.22 & 0.0602\\
 \hline
GC & MG & \bfseries 0.101 & 0.232 & \bfseries 0.0549 & \bfseries 0.0620 & \bfseries 0.0869 & 4.71 & 0.00966 & 0.15 & 0.097 & 4.76 & \bfseries 0.0494
\end{tabular}
}
\end{table*}

\begin{figure*}[h!]
 \includegraphics[width=1\textwidth]{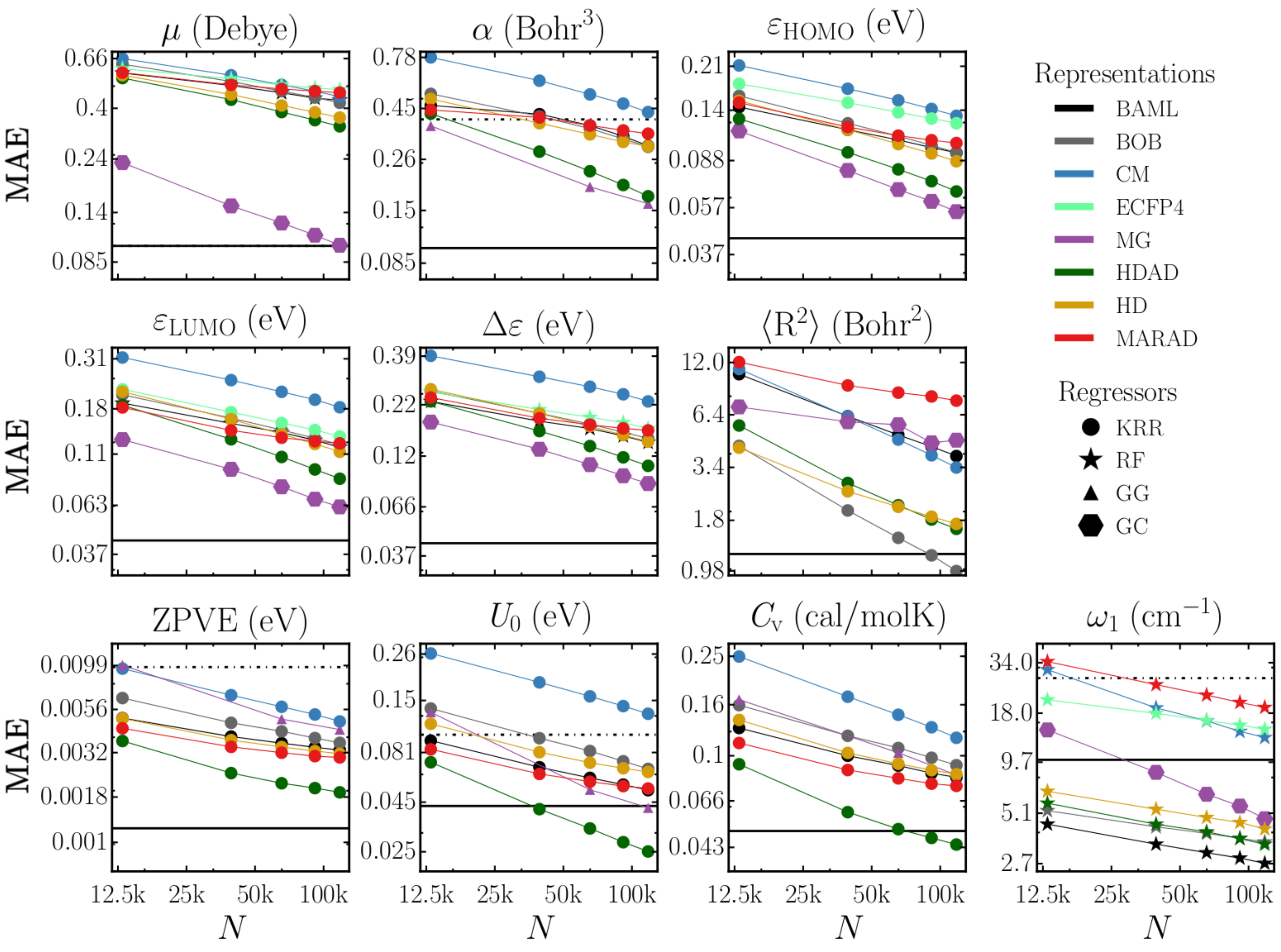}
 \caption{ \label{fig:learning_curves}
   Learning curves (mean absolute error (MAE) as a function of training set size $N$)
  for 10 properties of QM9 molecules, displaying the best regressor for each representation and property.
  Horizontal solid lines correspond to target accuracies, vertical dotted lines correspond to approximated B3LYP accuracies (unless off-chart), see also table \ref{tab:order}. Note that due to its poor performance ECFP4 results have been excluded for $\alpha$, $\langle \mathrm{R}^2 \rangle$, $\mathrm{ZPVE}$, $\mathrm{U}$ and $\mathrm{C}_{\mathrm{v}}$.  
  }
  
\end{figure*}

\subsection{Overview}
We present an overview of the most relevant numerical results in Table \ref{tab:order}. 
It contains the test errors for all combinations of regressors and representations and properties for
models trained on $\sim$117 k molecules. 
The best models for the respective properties are 
$\mu$ (MG/GC), 
$\alpha$ (MG/GG), 
$\varepsilon_{\mathrm{HOMO}}$ (MG/GC), 
$\varepsilon_{\mathrm{LUMO}}$ (MG/GC), 
$\Delta\varepsilon$ (MG/GC), 
$\langle\mathrm{R}^2\rangle$ (BOB/KRR),
$\mathrm{ZPVE}$ (HDAD/KRR),
$\mathit{U}_{\rm 0}$ (HDAD/KRR), 
$\mathit{C}_{\mathrm{v}}$ (HDAD/KRR),
and $\omega_1$ (BAML/RF).
We do not show results for the other three energies, $U(T=298K), H(T=298K), G(T=298K)$ since identical observations as for $U_0$ can be made.

Overall, NN and KRR regressors perform well for most properties. 
The ML out-of-sample errors outperform DFT accuracy and reach chemical (target) accuracy, both defined alongside in table \ref{tab:order}, for $\mu$ (GC), $\langle\mathrm{R}^2\rangle$ (BOB/KRR), $\mathit{U}_{\rm 0}$ (HDAD/KRR and MG/GG), $\mathit{C}_{\mathrm{v}}$ (HDAD/KRR), and $\omega_1$ (BAML/KRR, MG/GC, HDAD/KRR, BOB/KRR, HD/KRR and MG/GG). 
For the remaining properties ($\alpha$, $\varepsilon_{\mathrm{HOMO}}$, $\varepsilon_{\mathrm{LUMO}}$, $\Delta \epsilon$, and $\mathrm{ZPVE}$) the 
best models come within a factor 2 of target accuracy, while all outperforming hybrid DFT accuracy. 

In Fig.~\ref{fig:learning_curves} out-of-sample errors as a function of training set size (learning curves) are shown for all properties and representations with the best corresponding regressor. 
It is important to note that {\em all} models on display systematically improve
with training set size, exhibiting the typical linearly decaying behavior on a log-log plot~\cite{Muller1996,BAML}. 
Errors for most models shown decay with roughly the same slopes, 
indicating similar exponents in the power-law of error decay. 
Notable exceptions, i.e.~property models with considerably steeper learning curves, are
MG/GC for $\mu$, MG/GG and HDAD/KRR for $\alpha$, CM/KRR and BoB/KRR for $\langle {\rm R}^2\rangle$, HDAD/KRR and MG/GG for $U_{\rm 0}$, and MG/GG for $\omega_1$.
These results suggest that the specified representations capture particularly well the effective dimensionality of the corresponding porperty in chemical space. 


\subsection{Regressors}


Inspection of Table~\ref{tab:order} indicates that the regressors can roughly be ordered by performance, independent of property and representation: 
GC$>$GG$>$KRR$>$RF$>$BR$>$EN.
It is noteworthy how EN, BR, and RF regressors perform substantially worse than GC/GG/KRR. 
This can also be seen from the learning curves of all regressors presented in Figures S1 to S6 of the Supplementary Material.
The performance of BR and EN improves only slightly with increased training set size and even gets worse for some property/representation combinations. 
These two regressors also exhibit very similar learning curves and BR preforms only slightly better than EN for most combinations. 
The only clear exception to this rule is for ZPVE and $U_0$ together with HDAD, where BR preforms significantly better than EN. 
Also, BR and EN errors rapidly converge to a constant w.r.t. training set size for all representations and properties, except for HDAD, which is the only representation which has a noteworthy improvement with increased training set size for some properties. 
The constant learning rates are not surprising as (a) the number of free regression parameters in BR and EN is relatively small and does not grow with training set size, and as (b) the underlying model is a linear combination with small flexibility. 
This behavior implies error convergence already for relatively small training sets. 


RF performs poorly compared to GC, GG and KRR for all properties except for $\omega_1$, the highest lying fundamental vibrational frequency in each molecule. 
For this property RF yields an astounding performance with out-of-sample errors as small as single digit cm$^{-1}$. 
B3LYP achieves a mean absolute error of only 28 cm$^{-1}$ with respect to experiment~\cite{AngelaVibrationDFTerror2004}. 
The distribution of $\omega_1$, Figure 1 of reference \cite{sk}, suggests a simple reason for this: 
There are three distinct peaks which correspond to typical C-H, N-H and O-H stretch vibrations in increasing order. 
Therefore the principal learning task in this property is to detect if there is an OH group, and if not if there is an NH group. 
If neither group is present, CH will yield the vibration with the highest frequency. 
As such, this is essentially about classifying which bonds are present in the molecule.
RF works by fitting a decision tree to the target property. 
Each branch in the tree is based on an inequality of \emph{one} entry in the representation. 
RF should therefore be able to identify which bonds are present in a molecule, simply by looking 
at the entries in the each element pair, and/or triplet bin of the representations. 
For RF, a fractional importance can be assigned to each input feature (the sum of all importances is 1.0). 
Analyzing the importance of the bins in HDAD of the RF model reveals that the three bins with highest importance are: O-H placed at 0.961 \AA, N-H placed at at $1.01$ \AA\, and C-C-H at $3.138$ radians with an importance of 0.587, 0.305 and 0.064 respectively. 
These three first bins constitute  $\sim$96$\%$ of the prediction of $\omega_1$ and distances of the O-H and N-H bins are very similar to O-H and N-H bond lengths. 
C-C-H is placed on $\sim\pi$ radians which means that it has to correspond to a linear C-C-H (alkyne) chain which implies that the two carbons must be bonded by a triple bond (typically the C-H with the lowest pK$_a$ and the highest C-H stretch vibration). 

KRR performs remarkably well on average. For energetic properties, as well as $\langle R^2\rangle$, it yields the lowest overall errors in combination with HDAD and BoB, respectively. 
Its outstanding performance is not unsurprising in light of the multiple previous ML studies dealing with compositional as well as configurational spaces. 
The neural network flavors GC and GG, however, yield better performance on average, and the lowest errors for 
all electronic properties, i.e.~dipolemoment, polarizability, HOMO/LUMO eigenvalues and gaps.

\subsection{Representations}
As one would expect, HDAD contains more relevant information and thus it always outperforms HD when using KRR. 
Tests also showed that an HDA representation systematically yields errors in between HDAD and HD, and similar observations hold for BR and EN regressor. 
In the case of RF, however, we observe little difference between HDAD and HD, and HD can even yield slightly lower errors than HDAD. 
In our opinion, this is due to the additional bins of angles and dihedrals rather adding noise than signal. 
By contrast, the separation of distances, angles and dihedral angles into different bins is not a problem for the KRR methods because the kernels used are purely  distance based. This makes it possible for KRR to exploit the extra three- and four-body information in HDAD and to gain an advantage over HD. 
We note however that the remarkable performance of HDAD is possible despite its striking simplicity. 
As illustrated in Fig.~\ref{fig:hist_example} and discussed above, characteristic chemical behavior
can be directly obtained by human inspection of HDAD. As such, HDAD corresponds to a representation very much "Occam's razor style". Unfortunately, due to its discrete nature and its origin in sorting distances, HDAD will suffer from lack of differentiability, which might limit its applicability when modeling forces or other non-equilibrium properties.

MARAD, in principle containing the same information as HDA, performs similarly to BAML---yet, MARAD requires no prior
knowledge about the physics encoded in the universal force-field. 
BoB and CM, while previously considered state of the art, result in relatively poor performance except for $\langle R^2 \rangle$ where
BoB/KRR gives the lowest overall error. The simple bag of bonds, representing lists of all the pairwise Coulombic interactions
appears to coincide ideally with the corresponding electronic spread in the molecule. Due to its inherent lack of uniqueness, however, 
we believe that also this property must converge to a finite error in the limit of infinite training set size. 
ECFP4 produces out-of-sample errors on par or slightly better than CM/KRR for intensive properties ($\mu$, HOMO/LUMO eigenvalues and gap), 
however the model produces errors that are off-the-chart for all extensive properties ($\alpha$, $\langle\mathrm{R}^2\rangle$, $\mathrm{ZPVE}$, $U_0$ and $C_\mathrm{V}$).

\section{Conclusions}
We have benchmarked many combinations of regressors and representations on the same QM9 data set consisting of $\sim$131k organic molecules. 
For all properties, the best ML model prediction errors reach the accuracy of hybrid DFT with respect to experiment.
For 8 out of 13 distinct properties (atomization energies, heat-capacity, $\omega_1$, $\langle R^2\rangle$, $\mu$) 
out-of-sample errors reach levels on par with chemical accuracy, or better, using a training set size of $\sim$117k (90$\%$ of QM9 data set) molecules.
For the remaining properties $\alpha$, $\varepsilon_{\mathrm{HOMO}}$, $\varepsilon_{\mathrm{LUMO}}$, $\Delta \epsilon$, and $\mathrm{ZPVE}$,
errors of the best models come within a factor 2 of chemical accuracy.

Regressors EN, BR, and RF lead to rather high out-of-sample errors, while KRR and graph based NN regressors compete for the lowest errors. 
We have found that GC, GG, and KRR have best performance across {\em all} properties, except for the highest vibrational frequency for which RF performs best. 
There is no single representation and regressor combination that works best for all properties
(though forthcoming work with further improvements to the GG based models indicates best 
in class performance across all properties~\cite{NeuralMessagePassing}).
For intensive electronic properties ($\mu$, HOMO/LUMO eigenvalues and gap) we have found MG/NN to yield the highest predictive power,
while HDAD/KRR corresponds to the most accurate model for extensive properties ($\alpha$, $\langle\mathrm{R}^2\rangle$, $\mathrm{ZPVE}$, $U_0$ and $C_\mathrm{V}$). 
The latter point is remarkable when considering the simplicity of KRR, being just a linear expansion of property in training set, and 
HDAD, being just histograms of distances, angels, dihedrals.
Using BR and EN is not recommended if accuracy is of importance, both regressors perform worse across all properties. 
Apart from predicting highest fundamental vibrational frequency best, RF based models deliver rather unsatisfactory performance.
The ECPF4 based models have shown poor general performance in comparison to all other representations studied; 
it is not recommended for investigations of molecular properties.

We should caution the reader that all our results refer to equilibrium structures of a set of only $\sim$131 k organic molecules. 
While $\sim$131k molecules might seem sufficiently large to be representative, this number is dwarfed in comparison to chemical space, i.e. the space populated by all theoretically stable molecules, 
estimated to exceed 10$^{60}$ for medium sized organic molecules~\cite{ChemicalSpace}. 
Furthermore, ML models for predicting properties of molecules in non-equilibrium or strained  configurations might require substantially more training data. 
This point is also of relevance because some of the highly accurate models described herewithin (MG based) 
currently use bond based graph connectivity in addition to distance, raising questions about the applicability to reactive processes.

In summary, for the organic molecules studied, we have collected numerical evidence which suggests that the out-of-sample error of ML 
is consistently better than estimated DFT accuracy 
While this does not prove that ML models would reach same error levels if more accurate, explicitly electron correlated or experimental reference data was used, 
there are studies which suggest that relevant property  approximations do become smoother when using higher levels of theory~\cite{Goedecker2010,delta_learning}.
As such, our study indicates that future reference data sets for training modern state-of-the-art machine learning models will require levels of theory which go beyond hybrid DFT. 
Finally, we note that future work could deal with improving representations and regressors with the goal of reaching similar predictive power using less data.

\begin{acknowledgements}
The authors thank Dirk Bakowies for helpful comments, 
and Adrian Roitberg for pointing out an issue with the use of partial charges in the neural net models in an earlier version of this paper.
O.A.v.L. acknowledges support from the Swiss National Science foundation (No.~PP00P2\_138932, 310030\_160067), the research fund of the University of Basel, and from Google.
This material is based upon work supported by the Air Force Office of Scientific Research, Air Force Material Command, USAF under Award No. FA9550-15-1-0026. 
This research was partly supported by the NCCR MARVEL, 
funded by the Swiss National Science Foundation.
Some calculations were performed at sciCORE (http://scicore.unibas.ch/) scientific computing core facility at University of Basel.
\end{acknowledgements}

\bibliographystyle{apsrev4-1}
\bibliography{ref}

\begin{thebibliography}{65}%
\makeatletter
\providecommand \@ifxundefined [1]{%
 \@ifx{#1\undefined}
}%
\providecommand \@ifnum [1]{%
 \ifnum #1\expandafter \@firstoftwo
 \else \expandafter \@secondoftwo
 \fi
}%
\providecommand \@ifx [1]{%
 \ifx #1\expandafter \@firstoftwo
 \else \expandafter \@secondoftwo
 \fi
}%
\providecommand \natexlab [1]{#1}%
\providecommand \enquote  [1]{``#1''}%
\providecommand \bibnamefont  [1]{#1}%
\providecommand \bibfnamefont [1]{#1}%
\providecommand \citenamefont [1]{#1}%
\providecommand \href@noop [0]{\@secondoftwo}%
\providecommand \href [0]{\begingroup \@sanitize@url \@href}%
\providecommand \@href[1]{\@@startlink{#1}\@@href}%
\providecommand \@@href[1]{\endgroup#1\@@endlink}%
\providecommand \@sanitize@url [0]{\catcode `\\12\catcode `\$12\catcode
  `\&12\catcode `\#12\catcode `\^12\catcode `\_12\catcode `\%12\relax}%
\providecommand \@@startlink[1]{}%
\providecommand \@@endlink[0]{}%
\providecommand \url  [0]{\begingroup\@sanitize@url \@url }%
\providecommand \@url [1]{\endgroup\@href {#1}{\urlprefix }}%
\providecommand \urlprefix  [0]{URL }%
\providecommand \Eprint [0]{\href }%
\providecommand \doibase [0]{http://dx.doi.org/}%
\providecommand \selectlanguage [0]{\@gobble}%
\providecommand \bibinfo  [0]{\@secondoftwo}%
\providecommand \bibfield  [0]{\@secondoftwo}%
\providecommand \translation [1]{[#1]}%
\providecommand \BibitemOpen [0]{}%
\providecommand \bibitemStop [0]{}%
\providecommand \bibitemNoStop [0]{.\EOS\space}%
\providecommand \EOS [0]{\spacefactor3000\relax}%
\providecommand \BibitemShut  [1]{\csname bibitem#1\endcsname}%
\let\auto@bib@innerbib\@empty
\bibitem [{\citenamefont {Hohenberg}\ and\ \citenamefont {Kohn}(1964)}]{HK}%
  \BibitemOpen
  \bibfield  {author} {\bibinfo {author} {\bibfnamefont {P.}~\bibnamefont
  {Hohenberg}}\ and\ \bibinfo {author} {\bibfnamefont {W.}~\bibnamefont
  {Kohn}},\ }\href@noop {} {\bibfield  {journal} {\bibinfo  {journal} {Phys.
  Rev.}\ }\textbf {\bibinfo {volume} {136}},\ \bibinfo {pages} {B864} (\bibinfo
  {year} {1964})}\BibitemShut {NoStop}%
\bibitem [{\citenamefont {Kohn}\ and\ \citenamefont {Sham}(1965)}]{KS}%
  \BibitemOpen
  \bibfield  {author} {\bibinfo {author} {\bibfnamefont {W.}~\bibnamefont
  {Kohn}}\ and\ \bibinfo {author} {\bibfnamefont {L.~J.}\ \bibnamefont
  {Sham}},\ }\href@noop {} {\bibfield  {journal} {\bibinfo  {journal} {Phys.
  Rev.}\ }\textbf {\bibinfo {volume} {140}},\ \bibinfo {pages} {A1133}
  (\bibinfo {year} {1965})}\BibitemShut {NoStop}%
\bibitem [{\citenamefont {Burke}(2012)}]{BurkePerspectives_2012jcp}%
  \BibitemOpen
  \bibfield  {author} {\bibinfo {author} {\bibfnamefont {K.}~\bibnamefont
  {Burke}},\ }\href {\doibase http://dx.doi.org/10.1063/1.4704546} {\bibfield
  {journal} {\bibinfo  {journal} {J. Chem. Phys.}\ }\textbf {\bibinfo {volume}
  {136}},\ \bibinfo {pages} {150901} (\bibinfo {year} {2012})}\BibitemShut
  {NoStop}%
\bibitem [{\citenamefont {Koch}\ and\ \citenamefont
  {Holthausen}(2002)}]{ChemistsGuidetoDFT}%
  \BibitemOpen
  \bibfield  {author} {\bibinfo {author} {\bibfnamefont {W.}~\bibnamefont
  {Koch}}\ and\ \bibinfo {author} {\bibfnamefont {M.~C.}\ \bibnamefont
  {Holthausen}},\ }\href@noop {} {\emph {\bibinfo {title} {A Chemist's Guide to
  Density Functional Theory}}}\ (\bibinfo  {publisher} {Wiley-VCH},\ \bibinfo
  {year} {2002})\BibitemShut {NoStop}%
\bibitem [{\citenamefont {Cohen}\ \emph {et~al.}(2012)\citenamefont {Cohen},
  \citenamefont {Mori-S\'anchez},\ and\ \citenamefont {Yang}}]{CohenYang2012}%
  \BibitemOpen
  \bibfield  {author} {\bibinfo {author} {\bibfnamefont {A.~J.}\ \bibnamefont
  {Cohen}}, \bibinfo {author} {\bibfnamefont {P.}~\bibnamefont
  {Mori-S\'anchez}}, \ and\ \bibinfo {author} {\bibfnamefont {W.}~\bibnamefont
  {Yang}},\ }\href {\doibase 10.1021/cr200107z} {\bibfield  {journal} {\bibinfo
   {journal} {Chem. Rev.}\ }\textbf {\bibinfo {volume} {112}},\ \bibinfo
  {pages} {289} (\bibinfo {year} {2012})}\BibitemShut {NoStop}%
\bibitem [{\citenamefont {Plata}\ and\ \citenamefont
  {Singleton}(2015)}]{SingletonDFTflawed2015}%
  \BibitemOpen
  \bibfield  {author} {\bibinfo {author} {\bibfnamefont {R.~E.}\ \bibnamefont
  {Plata}}\ and\ \bibinfo {author} {\bibfnamefont {D.~A.}\ \bibnamefont
  {Singleton}},\ }\href {\doibase 10.1021/ja5111392} {\bibfield  {journal}
  {\bibinfo  {journal} {J. Am. Chem. Soc.}\ }\textbf {\bibinfo {volume}
  {137}},\ \bibinfo {pages} {3811} (\bibinfo {year} {2015})}\BibitemShut
  {NoStop}%
\bibitem [{\citenamefont {Medvedev}\ \emph {et~al.}(2017)\citenamefont
  {Medvedev}, \citenamefont {Bushmarinov}, \citenamefont {Sun}, \citenamefont
  {Perdew},\ and\ \citenamefont {Lyssenko}}]{PerdewDFTScience2017}%
  \BibitemOpen
  \bibfield  {author} {\bibinfo {author} {\bibfnamefont {M.~G.}\ \bibnamefont
  {Medvedev}}, \bibinfo {author} {\bibfnamefont {I.~S.}\ \bibnamefont
  {Bushmarinov}}, \bibinfo {author} {\bibfnamefont {J.}~\bibnamefont {Sun}},
  \bibinfo {author} {\bibfnamefont {J.~P.}\ \bibnamefont {Perdew}}, \ and\
  \bibinfo {author} {\bibfnamefont {K.~A.}\ \bibnamefont {Lyssenko}},\ }\href
  {\doibase 10.1126/science.aah5975} {\bibfield  {journal} {\bibinfo  {journal}
  {Science}\ }\textbf {\bibinfo {volume} {355}},\ \bibinfo {pages} {49}
  (\bibinfo {year} {2017})}\BibitemShut {NoStop}%
\bibitem [{\citenamefont {Ramakrishnan}\ \emph {et~al.}(2015)\citenamefont
  {Ramakrishnan}, \citenamefont {Dral}, \citenamefont {Rupp},\ and\
  \citenamefont {von Lilienfeld}}]{delta_learning}%
  \BibitemOpen
  \bibfield  {author} {\bibinfo {author} {\bibfnamefont {R.}~\bibnamefont
  {Ramakrishnan}}, \bibinfo {author} {\bibfnamefont {P.~O.}\ \bibnamefont
  {Dral}}, \bibinfo {author} {\bibfnamefont {M.}~\bibnamefont {Rupp}}, \ and\
  \bibinfo {author} {\bibfnamefont {O.~A.}\ \bibnamefont {von Lilienfeld}},\
  }\href {\doibase 10.1021/acs.jctc.5b00099} {\bibfield  {journal} {\bibinfo
  {journal} {J. Chem. Theory Comput.}\ }\textbf {\bibinfo {volume} {11}},\
  \bibinfo {pages} {2087} (\bibinfo {year} {2015})}\BibitemShut {NoStop}%
\bibitem [{\citenamefont {Ramakrishnan}\ \emph {et~al.}(2014)\citenamefont
  {Ramakrishnan}, \citenamefont {Dral}, \citenamefont {Rupp},\ and\
  \citenamefont {von Lilienfeld}}]{gdb9}%
  \BibitemOpen
  \bibfield  {author} {\bibinfo {author} {\bibfnamefont {R.}~\bibnamefont
  {Ramakrishnan}}, \bibinfo {author} {\bibfnamefont {P.~O.}\ \bibnamefont
  {Dral}}, \bibinfo {author} {\bibfnamefont {M.}~\bibnamefont {Rupp}}, \ and\
  \bibinfo {author} {\bibfnamefont {O.~A.}\ \bibnamefont {von Lilienfeld}},\
  }\href {http://dx.doi.org/10.1038/sdata.2014.22} {\bibfield  {journal}
  {\bibinfo  {journal} {Sci. Data}\ }\textbf {\bibinfo {volume} {1}} (\bibinfo
  {year} {2014})}\BibitemShut {NoStop}%
\bibitem [{\citenamefont {Ruddigkeit}\ \emph {et~al.}(2012)\citenamefont
  {Ruddigkeit}, \citenamefont {van Deursen}, \citenamefont {Blum},\ and\
  \citenamefont {Reymond}}]{gdb17}%
  \BibitemOpen
  \bibfield  {author} {\bibinfo {author} {\bibfnamefont {L.}~\bibnamefont
  {Ruddigkeit}}, \bibinfo {author} {\bibfnamefont {R.}~\bibnamefont {van
  Deursen}}, \bibinfo {author} {\bibfnamefont {L.~C.}\ \bibnamefont {Blum}}, \
  and\ \bibinfo {author} {\bibfnamefont {J.-L.}\ \bibnamefont {Reymond}},\
  }\href {\doibase 10.1021/ci300415d} {\bibfield  {journal} {\bibinfo
  {journal} {J. Chem. Inf. Model.}\ }\textbf {\bibinfo {volume} {52}},\
  \bibinfo {pages} {2864} (\bibinfo {year} {2012})}\BibitemShut {NoStop}%
\bibitem [{\citenamefont {Huang}\ and\ \citenamefont {von
  Lilienfeld}(2016)}]{BAML}%
  \BibitemOpen
  \bibfield  {author} {\bibinfo {author} {\bibfnamefont {B.}~\bibnamefont
  {Huang}}\ and\ \bibinfo {author} {\bibfnamefont {O.~A.}\ \bibnamefont {von
  Lilienfeld}},\ }\href {\doibase 10.1063/1.4964627} {\bibfield  {journal}
  {\bibinfo  {journal} {J. Chem. Phys}\ }\textbf {\bibinfo {volume} {145}},\
  \bibinfo {eid} {161102} (\bibinfo {year} {2016})}\BibitemShut {NoStop}%
\bibitem [{\citenamefont {Hansen}\ \emph {et~al.}(2015)\citenamefont {Hansen},
  \citenamefont {Biegler}, \citenamefont {Ramakrishnan}, \citenamefont
  {Pronobis}, \citenamefont {von Lilienfeld}, \citenamefont {M{\"u}ller},\ and\
  \citenamefont {Tkatchenko}}]{bob}%
  \BibitemOpen
  \bibfield  {author} {\bibinfo {author} {\bibfnamefont {K.}~\bibnamefont
  {Hansen}}, \bibinfo {author} {\bibfnamefont {F.}~\bibnamefont {Biegler}},
  \bibinfo {author} {\bibfnamefont {R.}~\bibnamefont {Ramakrishnan}}, \bibinfo
  {author} {\bibfnamefont {W.}~\bibnamefont {Pronobis}}, \bibinfo {author}
  {\bibfnamefont {O.~A.}\ \bibnamefont {von Lilienfeld}}, \bibinfo {author}
  {\bibfnamefont {K.-R.}\ \bibnamefont {M{\"u}ller}}, \ and\ \bibinfo {author}
  {\bibfnamefont {A.}~\bibnamefont {Tkatchenko}},\ }\href {\doibase
  10.1021/acs.jpclett.5b00831} {\bibfield  {journal} {\bibinfo  {journal} {J.
  Phys. Chem. Lett.}\ }\textbf {\bibinfo {volume} {6}},\ \bibinfo {pages}
  {2326} (\bibinfo {year} {2015})}\BibitemShut {NoStop}%
\bibitem [{\citenamefont {Ramakrishnan}\ and\ \citenamefont {von
  Lilienfeld}(2015)}]{sk}%
  \BibitemOpen
  \bibfield  {author} {\bibinfo {author} {\bibfnamefont {R.}~\bibnamefont
  {Ramakrishnan}}\ and\ \bibinfo {author} {\bibfnamefont {O.~A.}\ \bibnamefont
  {von Lilienfeld}},\ }\href {\doibase doi:10.2533/chimia.2015.182} {\bibfield
  {journal} {\bibinfo  {journal} {chimia}\ }\textbf {\bibinfo {volume} {69}},\
  \bibinfo {pages} {182} (\bibinfo {year} {2015})}\BibitemShut {NoStop}%
\bibitem [{\citenamefont {Rupp}\ \emph {et~al.}(2012)\citenamefont {Rupp},
  \citenamefont {Tkatchenko},\ and\ \citenamefont {von Lilienfeld}}]{ML0}%
  \BibitemOpen
  \bibfield  {author} {\bibinfo {author} {\bibfnamefont {M.}~\bibnamefont
  {Rupp}}, \bibinfo {author} {\bibfnamefont {K.-R.}\ \bibnamefont {Tkatchenko},
  \bibfnamefont {Alexandre haand~M{\"u}ller}}, \ and\ \bibinfo {author}
  {\bibfnamefont {O.~A.}\ \bibnamefont {von Lilienfeld}},\ }\href
  {http://dx.doi.org/10.1103/PhysRevLett.108.058301} {\bibfield  {journal}
  {\bibinfo  {journal} {Phys. Rev. Lett.}\ }\textbf {\bibinfo {volume} {108}},\
  \bibinfo {pages} {058301} (\bibinfo {year} {2012})}\BibitemShut {NoStop}%
\bibitem [{\citenamefont {Barker}\ \emph {et~al.}(2016)\citenamefont {Barker},
  \citenamefont {Bulin}, \citenamefont {Hamaekers},\ and\ \citenamefont
  {Mathias}}]{barker2016localized}%
  \BibitemOpen
  \bibfield  {author} {\bibinfo {author} {\bibfnamefont {J.}~\bibnamefont
  {Barker}}, \bibinfo {author} {\bibfnamefont {J.}~\bibnamefont {Bulin}},
  \bibinfo {author} {\bibfnamefont {J.}~\bibnamefont {Hamaekers}}, \ and\
  \bibinfo {author} {\bibfnamefont {S.}~\bibnamefont {Mathias}},\ }\href@noop
  {} {\bibfield  {journal} {\bibinfo  {journal} {arXiv preprint
  arXiv:1611.05126}\ } (\bibinfo {year} {2016})}\BibitemShut {NoStop}%
\bibitem [{\citenamefont {von Lilienfeld}\ \emph {et~al.}(2015)\citenamefont
  {von Lilienfeld}, \citenamefont {Ramakrishnan}, \citenamefont {Rupp},\ and\
  \citenamefont {Knoll}}]{OAvL_FRD}%
  \BibitemOpen
  \bibfield  {author} {\bibinfo {author} {\bibfnamefont {O.~A.}\ \bibnamefont
  {von Lilienfeld}}, \bibinfo {author} {\bibfnamefont {R.}~\bibnamefont
  {Ramakrishnan}}, \bibinfo {author} {\bibfnamefont {M.}~\bibnamefont {Rupp}},
  \ and\ \bibinfo {author} {\bibfnamefont {A.}~\bibnamefont {Knoll}},\ }\href
  {\doibase 10.1002/qua.24912} {\bibfield  {journal} {\bibinfo  {journal} {Int.
  J. Quantum}\ }\textbf {\bibinfo {volume} {115}},\ \bibinfo {pages} {1084}
  (\bibinfo {year} {2015})}\BibitemShut {NoStop}%
\bibitem [{\citenamefont {Huan}\ \emph {et~al.}(2015)\citenamefont {Huan},
  \citenamefont {Mannodi-Kanakkithodi},\ and\ \citenamefont
  {Ramprasad}}]{Motif-Descriptor}%
  \BibitemOpen
  \bibfield  {author} {\bibinfo {author} {\bibfnamefont {T.~D.}\ \bibnamefont
  {Huan}}, \bibinfo {author} {\bibfnamefont {A.}~\bibnamefont
  {Mannodi-Kanakkithodi}}, \ and\ \bibinfo {author} {\bibfnamefont
  {R.}~\bibnamefont {Ramprasad}},\ }\href {\doibase 10.1103/PhysRevB.92.014106}
  {\bibfield  {journal} {\bibinfo  {journal} {Phys. Rev. B}\ }\textbf {\bibinfo
  {volume} {92}},\ \bibinfo {pages} {014106} (\bibinfo {year}
  {2015})}\BibitemShut {NoStop}%
\bibitem [{\citenamefont {Bart\'ok}\ \emph {et~al.}(2013)\citenamefont
  {Bart\'ok}, \citenamefont {Kondor},\ and\ \citenamefont
  {Cs\'anyi}}]{SOAP_original}%
  \BibitemOpen
  \bibfield  {author} {\bibinfo {author} {\bibfnamefont {A.~P.}\ \bibnamefont
  {Bart\'ok}}, \bibinfo {author} {\bibfnamefont {R.}~\bibnamefont {Kondor}}, \
  and\ \bibinfo {author} {\bibfnamefont {G.}~\bibnamefont {Cs\'anyi}},\ }\href
  {\doibase 10.1103/PhysRevB.87.184115} {\bibfield  {journal} {\bibinfo
  {journal} {Phys. Rev. B}\ }\textbf {\bibinfo {volume} {87}},\ \bibinfo
  {pages} {184115} (\bibinfo {year} {2013})}\BibitemShut {NoStop}%
\bibitem [{\citenamefont {De}\ \emph {et~al.}(2016)\citenamefont {De},
  \citenamefont {Bart\'ok}, \citenamefont {Cs\'anyi},\ and\ \citenamefont
  {Ceriotti}}]{SOAP_apl}%
  \BibitemOpen
  \bibfield  {author} {\bibinfo {author} {\bibfnamefont {S.}~\bibnamefont
  {De}}, \bibinfo {author} {\bibfnamefont {A.~P.}\ \bibnamefont {Bart\'ok}},
  \bibinfo {author} {\bibfnamefont {G.}~\bibnamefont {Cs\'anyi}}, \ and\
  \bibinfo {author} {\bibfnamefont {M.}~\bibnamefont {Ceriotti}},\ }\href
  {\doibase 10.1039/C6CP00415F} {\bibfield  {journal} {\bibinfo  {journal}
  {Phys. Chem. Chem. Phys.}\ }\textbf {\bibinfo {volume} {18}},\ \bibinfo
  {pages} {13754} (\bibinfo {year} {2016})}\BibitemShut {NoStop}%
\bibitem [{\citenamefont {Collins}\ \emph {et~al.}(2016)\citenamefont
  {Collins}, \citenamefont {Gordon}, \citenamefont {von Lilienfeld},\ and\
  \citenamefont {Yaron}}]{const_size_desc}%
  \BibitemOpen
  \bibfield  {author} {\bibinfo {author} {\bibfnamefont {C.~R.}\ \bibnamefont
  {Collins}}, \bibinfo {author} {\bibfnamefont {G.~J.}\ \bibnamefont {Gordon}},
  \bibinfo {author} {\bibfnamefont {O.~A.}\ \bibnamefont {von Lilienfeld}}, \
  and\ \bibinfo {author} {\bibfnamefont {D.~J.}\ \bibnamefont {Yaron}},\
  }\href@noop {} {\bibfield  {journal} {\bibinfo  {journal}
  {https://arxiv.org/abs/1701.06649}\ } (\bibinfo {year} {2016})}\BibitemShut
  {NoStop}%
\bibitem [{\citenamefont {Smith}\ \emph {et~al.}(2017)\citenamefont {Smith},
  \citenamefont {Isayev},\ and\ \citenamefont
  {Roitberg}}]{ANI_IsayevRoitberg2017}%
  \BibitemOpen
  \bibfield  {author} {\bibinfo {author} {\bibfnamefont {J.~S.}\ \bibnamefont
  {Smith}}, \bibinfo {author} {\bibfnamefont {O.}~\bibnamefont {Isayev}}, \
  and\ \bibinfo {author} {\bibfnamefont {A.~E.}\ \bibnamefont {Roitberg}},\
  }\href {\doibase 10.1039/C6SC05720A} {\bibfield  {journal} {\bibinfo
  {journal} {Chem. Sci.}\ ,\ } (\bibinfo {year} {2017})}\BibitemShut {NoStop}%
\bibitem [{\citenamefont {Sch{\"u}tt}\ \emph {et~al.}(2017)\citenamefont
  {Sch{\"u}tt}, \citenamefont {Arbabzadah}, \citenamefont {Chmiela},
  \citenamefont {M{\"u}ller},\ and\ \citenamefont
  {Tkatchenko}}]{DeepTensorNN_2017}%
  \BibitemOpen
  \bibfield  {author} {\bibinfo {author} {\bibfnamefont {K.~T.}\ \bibnamefont
  {Sch{\"u}tt}}, \bibinfo {author} {\bibfnamefont {F.}~\bibnamefont
  {Arbabzadah}}, \bibinfo {author} {\bibfnamefont {S.}~\bibnamefont {Chmiela}},
  \bibinfo {author} {\bibfnamefont {K.~R.}\ \bibnamefont {M{\"u}ller}}, \ and\
  \bibinfo {author} {\bibfnamefont {A.}~\bibnamefont {Tkatchenko}},\
  }\href@noop {} {\bibfield  {journal} {\bibinfo  {journal} {Nat. Commun.}\
  }\textbf {\bibinfo {volume} {8}},\ \bibinfo {pages} {13890} (\bibinfo {year}
  {2017})}\BibitemShut {NoStop}%
\bibitem [{\citenamefont {Dral}\ \emph {et~al.}(2015)\citenamefont {Dral},
  \citenamefont {von Lilienfeld},\ and\ \citenamefont {Thiel}}]{SEMIEMP_ML1}%
  \BibitemOpen
  \bibfield  {author} {\bibinfo {author} {\bibfnamefont {P.~O.}\ \bibnamefont
  {Dral}}, \bibinfo {author} {\bibfnamefont {O.~A.}\ \bibnamefont {von
  Lilienfeld}}, \ and\ \bibinfo {author} {\bibfnamefont {W.}~\bibnamefont
  {Thiel}},\ }\href {\doibase 10.1021/acs.jctc.5b00141} {\bibfield  {journal}
  {\bibinfo  {journal} {J. Chem. Theory}\ }\textbf {\bibinfo {volume} {11}},\
  \bibinfo {pages} {2120} (\bibinfo {year} {2015})}\BibitemShut {NoStop}%
\bibitem [{\citenamefont {Weber}\ and\ \citenamefont {Thiel}(2000)}]{OM2}%
  \BibitemOpen
  \bibfield  {author} {\bibinfo {author} {\bibfnamefont {W.}~\bibnamefont
  {Weber}}\ and\ \bibinfo {author} {\bibfnamefont {W.}~\bibnamefont {Thiel}},\
  }\href {\doibase 10.1007/s002149900083} {\bibfield  {journal} {\bibinfo
  {journal} {Theor. Chem. Acc.}\ }\textbf {\bibinfo {volume} {103}},\ \bibinfo
  {pages} {495} (\bibinfo {year} {2000})}\BibitemShut {NoStop}%
\bibitem [{\citenamefont {Dral}\ \emph {et~al.}(2016)\citenamefont {Dral},
  \citenamefont {Wu}, \citenamefont {Sp{\"o}rkel}, \citenamefont {Koslowski},\
  and\ \citenamefont {Thiel}}]{SEMIEMP_ML2}%
  \BibitemOpen
  \bibfield  {author} {\bibinfo {author} {\bibfnamefont {P.~O.}\ \bibnamefont
  {Dral}}, \bibinfo {author} {\bibfnamefont {X.}~\bibnamefont {Wu}}, \bibinfo
  {author} {\bibfnamefont {L.}~\bibnamefont {Sp{\"o}rkel}}, \bibinfo {author}
  {\bibfnamefont {A.}~\bibnamefont {Koslowski}}, \ and\ \bibinfo {author}
  {\bibfnamefont {W.}~\bibnamefont {Thiel}},\ }\href {\doibase
  10.1021/acs.jctc.5b01047} {\bibfield  {journal} {\bibinfo  {journal} {J.
  Chem. Theory}\ }\textbf {\bibinfo {volume} {12}},\ \bibinfo {pages} {1097}
  (\bibinfo {year} {2016})}\BibitemShut {NoStop}%
\bibitem [{\citenamefont {Hansen}\ \emph {et~al.}(2013)\citenamefont {Hansen},
  \citenamefont {Montavon}, \citenamefont {Biegler}, \citenamefont {Fazli},
  \citenamefont {Rupp}, \citenamefont {Scheffler}, \citenamefont {von
  Lilienfeld}, \citenamefont {Tkatchenko},\ and\ \citenamefont
  {M{\"u}ller}}]{AssessmentMLJCTC2013}%
  \BibitemOpen
  \bibfield  {author} {\bibinfo {author} {\bibfnamefont {K.}~\bibnamefont
  {Hansen}}, \bibinfo {author} {\bibfnamefont {G.}~\bibnamefont {Montavon}},
  \bibinfo {author} {\bibfnamefont {F.}~\bibnamefont {Biegler}}, \bibinfo
  {author} {\bibfnamefont {S.}~\bibnamefont {Fazli}}, \bibinfo {author}
  {\bibfnamefont {M.}~\bibnamefont {Rupp}}, \bibinfo {author} {\bibfnamefont
  {M.}~\bibnamefont {Scheffler}}, \bibinfo {author} {\bibfnamefont {O.~A.}\
  \bibnamefont {von Lilienfeld}}, \bibinfo {author} {\bibfnamefont
  {A.}~\bibnamefont {Tkatchenko}}, \ and\ \bibinfo {author} {\bibfnamefont
  {K.-R.}\ \bibnamefont {M{\"u}ller}},\ }\href {\doibase 10.1021/ct400195d}
  {\bibfield  {journal} {\bibinfo  {journal} {J. Chem. Theory Comput.}\
  }\textbf {\bibinfo {volume} {9}},\ \bibinfo {pages} {3404} (\bibinfo {year}
  {2013})}\BibitemShut {NoStop}%
\bibitem [{\citenamefont {Kearnes}\ \emph {et~al.}(2016)\citenamefont
  {Kearnes}, \citenamefont {McCloskey}, \citenamefont {Berndl}, \citenamefont
  {Pande},\ and\ \citenamefont {Riley}}]{kearnes2016molecular}%
  \BibitemOpen
  \bibfield  {author} {\bibinfo {author} {\bibfnamefont {S.}~\bibnamefont
  {Kearnes}}, \bibinfo {author} {\bibfnamefont {K.}~\bibnamefont {McCloskey}},
  \bibinfo {author} {\bibfnamefont {M.}~\bibnamefont {Berndl}}, \bibinfo
  {author} {\bibfnamefont {V.}~\bibnamefont {Pande}}, \ and\ \bibinfo {author}
  {\bibfnamefont {P.}~\bibnamefont {Riley}},\ }\href@noop {} {\bibfield
  {journal} {\bibinfo  {journal} {J. Comput. Aided Mol. Des.}\ }\textbf
  {\bibinfo {volume} {30}},\ \bibinfo {pages} {595} (\bibinfo {year}
  {2016})}\BibitemShut {NoStop}%
\bibitem [{\citenamefont {Li}\ \emph {et~al.}(2016)\citenamefont {Li},
  \citenamefont {Tarlow}, \citenamefont {Brockschmidt},\ and\ \citenamefont
  {Zemel}}]{yujia}%
  \BibitemOpen
  \bibfield  {author} {\bibinfo {author} {\bibfnamefont {Y.}~\bibnamefont
  {Li}}, \bibinfo {author} {\bibfnamefont {D.}~\bibnamefont {Tarlow}}, \bibinfo
  {author} {\bibfnamefont {M.}~\bibnamefont {Brockschmidt}}, \ and\ \bibinfo
  {author} {\bibfnamefont {R.}~\bibnamefont {Zemel}},\ }\href@noop {}
  {\bibfield  {journal} {\bibinfo  {journal} {ICLR}\ } (\bibinfo {year}
  {2016})}\BibitemShut {NoStop}%
\bibitem [{\citenamefont {Hickey}\ and\ \citenamefont
  {Rowley}(2014)}]{hickey2014benchmarking}%
  \BibitemOpen
  \bibfield  {author} {\bibinfo {author} {\bibfnamefont {A.~L.}\ \bibnamefont
  {Hickey}}\ and\ \bibinfo {author} {\bibfnamefont {C.~N.}\ \bibnamefont
  {Rowley}},\ }\href@noop {} {\bibfield  {journal} {\bibinfo  {journal} {J.
  Phys. Chem. A}\ }\textbf {\bibinfo {volume} {118}},\ \bibinfo {pages} {3678}
  (\bibinfo {year} {2014})}\BibitemShut {NoStop}%
\bibitem [{\citenamefont {Stowasser}\ and\ \citenamefont
  {Hoffmann}(1999)}]{MeaningOfKSorbitals}%
  \BibitemOpen
  \bibfield  {author} {\bibinfo {author} {\bibfnamefont {R.}~\bibnamefont
  {Stowasser}}\ and\ \bibinfo {author} {\bibfnamefont {R.}~\bibnamefont
  {Hoffmann}},\ }\href@noop {} {\bibfield  {journal} {\bibinfo  {journal} {J.
  Am. Chem. Soc.}\ }\textbf {\bibinfo {volume} {121}},\ \bibinfo {pages} {3414}
  (\bibinfo {year} {1999})}\BibitemShut {NoStop}%
\bibitem [{nis(2016)}]{nist}%
  \BibitemOpen
  \href@noop {} {\bibfield  {journal} {\bibinfo  {journal} {NIST computational
  chemistry comparison and benchmark database, Editor: Russell D. Johnson III,
  http://cccbdb.nist.gov/}\ } (\bibinfo {year} {2016})}\BibitemShut {NoStop}%
\bibitem [{\citenamefont {Parr}\ and\ \citenamefont {Yang}(1989)}]{parryang}%
  \BibitemOpen
  \bibfield  {author} {\bibinfo {author} {\bibfnamefont {R.~G.}\ \bibnamefont
  {Parr}}\ and\ \bibinfo {author} {\bibfnamefont {W.}~\bibnamefont {Yang}},\
  }\href@noop {} {\emph {\bibinfo {title} {Density functional theory of atoms
  and molecules}}}\ (\bibinfo  {publisher} {Oxford Science Publications},\
  \bibinfo {year} {1989})\BibitemShut {NoStop}%
\bibitem [{\citenamefont {Sinha}\ \emph {et~al.}(2004)\citenamefont {Sinha},
  \citenamefont {Boesch}, \citenamefont {Gu}, \citenamefont {Wheeler},\ and\
  \citenamefont {Wilson}}]{AngelaVibrationDFTerror2004}%
  \BibitemOpen
  \bibfield  {author} {\bibinfo {author} {\bibfnamefont {P.}~\bibnamefont
  {Sinha}}, \bibinfo {author} {\bibfnamefont {S.~E.}\ \bibnamefont {Boesch}},
  \bibinfo {author} {\bibfnamefont {C.}~\bibnamefont {Gu}}, \bibinfo {author}
  {\bibfnamefont {R.~A.}\ \bibnamefont {Wheeler}}, \ and\ \bibinfo {author}
  {\bibfnamefont {A.~K.}\ \bibnamefont {Wilson}},\ }\href {\doibase
  10.1021/jp048233q} {\bibfield  {journal} {\bibinfo  {journal} {J. Phys. Chem.
  A}\ }\textbf {\bibinfo {volume} {108}},\ \bibinfo {pages} {9213} (\bibinfo
  {year} {2004})}\BibitemShut {NoStop}%
\bibitem [{\citenamefont {Geary}(1935)}]{Geary1935MAE_RMSE_RATIO}%
  \BibitemOpen
  \bibfield  {author} {\bibinfo {author} {\bibfnamefont {R.~C.}\ \bibnamefont
  {Geary}},\ }\href@noop {} {\bibfield  {journal} {\bibinfo  {journal}
  {Biometrika}\ }\textbf {\bibinfo {volume} {27}},\ \bibinfo {pages} {310}
  (\bibinfo {year} {1935})}\BibitemShut {NoStop}%
\bibitem [{\citenamefont {Curtiss}\ \emph {et~al.}(1997)\citenamefont
  {Curtiss}, \citenamefont {Raghavachari}, \citenamefont {Redfern},\ and\
  \citenamefont {Pople}}]{JhonDFTEnthalpies1997}%
  \BibitemOpen
  \bibfield  {author} {\bibinfo {author} {\bibfnamefont {L.~A.}\ \bibnamefont
  {Curtiss}}, \bibinfo {author} {\bibfnamefont {K.}~\bibnamefont
  {Raghavachari}}, \bibinfo {author} {\bibfnamefont {P.~C.}\ \bibnamefont
  {Redfern}}, \ and\ \bibinfo {author} {\bibfnamefont {J.~A.}\ \bibnamefont
  {Pople}},\ }\href {\doibase 10.1063/1.473182} {\bibfield  {journal} {\bibinfo
   {journal} {J. Chem. Phys.}\ }\textbf {\bibinfo {volume} {106}},\ \bibinfo
  {pages} {1063} (\bibinfo {year} {1997})}\BibitemShut {NoStop}%
\bibitem [{\citenamefont {DeTar}(2007)}]{DeLosCv2007}%
  \BibitemOpen
  \bibfield  {author} {\bibinfo {author} {\bibfnamefont {D.~F.}\ \bibnamefont
  {DeTar}},\ }\href {\doibase 10.1021/jp066312r} {\bibfield  {journal}
  {\bibinfo  {journal} {J. Phys. Chem. A}\ }\textbf {\bibinfo {volume} {111}},\
  \bibinfo {pages} {4464} (\bibinfo {year} {2007})}\BibitemShut {NoStop}%
\bibitem [{\citenamefont {Rappe}\ \emph {et~al.}(1992)\citenamefont {Rappe},
  \citenamefont {Casewit}, \citenamefont {Colwell}, \citenamefont {III},\ and\
  \citenamefont {Skiff}}]{uff}%
  \BibitemOpen
  \bibfield  {author} {\bibinfo {author} {\bibfnamefont {A.~K.}\ \bibnamefont
  {Rappe}}, \bibinfo {author} {\bibfnamefont {C.~J.}\ \bibnamefont {Casewit}},
  \bibinfo {author} {\bibfnamefont {K.~S.}\ \bibnamefont {Colwell}}, \bibinfo
  {author} {\bibfnamefont {W.~A.~G.}\ \bibnamefont {III}}, \ and\ \bibinfo
  {author} {\bibfnamefont {W.~M.}\ \bibnamefont {Skiff}},\ }\href {\doibase
  10.1021/ja00051a040} {\bibfield  {journal} {\bibinfo  {journal} {J. Am. Chem.
  Soc.}\ }\textbf {\bibinfo {volume} {114}},\ \bibinfo {pages} {10024}
  (\bibinfo {year} {1992})}\BibitemShut {NoStop}%
\bibitem [{\citenamefont {Rogers}\ and\ \citenamefont
  {Hahn}(2010)}]{rogers2010extended}%
  \BibitemOpen
  \bibfield  {author} {\bibinfo {author} {\bibfnamefont {D.}~\bibnamefont
  {Rogers}}\ and\ \bibinfo {author} {\bibfnamefont {M.}~\bibnamefont {Hahn}},\
  }\href@noop {} {\bibfield  {journal} {\bibinfo  {journal} {J. Chem. Inf.
  Model.}\ }\textbf {\bibinfo {volume} {50}},\ \bibinfo {pages} {742} (\bibinfo
  {year} {2010})}\BibitemShut {NoStop}%
\bibitem [{\citenamefont {Besnard}\ \emph {et~al.}(2012)\citenamefont
  {Besnard}, \citenamefont {Ruda}, \citenamefont {Setola}, \citenamefont
  {Abecassis}, \citenamefont {Rodriguiz}, \citenamefont {Huang}, \citenamefont
  {Norval}, \citenamefont {Sassano}, \citenamefont {Shin}, \citenamefont
  {Webster} \emph {et~al.}}]{besnard2012automated}%
  \BibitemOpen
  \bibfield  {author} {\bibinfo {author} {\bibfnamefont {J.}~\bibnamefont
  {Besnard}}, \bibinfo {author} {\bibfnamefont {G.~F.}\ \bibnamefont {Ruda}},
  \bibinfo {author} {\bibfnamefont {V.}~\bibnamefont {Setola}}, \bibinfo
  {author} {\bibfnamefont {K.}~\bibnamefont {Abecassis}}, \bibinfo {author}
  {\bibfnamefont {R.~M.}\ \bibnamefont {Rodriguiz}}, \bibinfo {author}
  {\bibfnamefont {X.-P.}\ \bibnamefont {Huang}}, \bibinfo {author}
  {\bibfnamefont {S.}~\bibnamefont {Norval}}, \bibinfo {author} {\bibfnamefont
  {M.~F.}\ \bibnamefont {Sassano}}, \bibinfo {author} {\bibfnamefont {A.~I.}\
  \bibnamefont {Shin}}, \bibinfo {author} {\bibfnamefont {L.~A.}\ \bibnamefont
  {Webster}},  \emph {et~al.},\ }\href@noop {} {\bibfield  {journal} {\bibinfo
  {journal} {Nature}\ }\textbf {\bibinfo {volume} {492}},\ \bibinfo {pages}
  {215} (\bibinfo {year} {2012})}\BibitemShut {NoStop}%
\bibitem [{\citenamefont {Lounkine}\ \emph {et~al.}(2012)\citenamefont
  {Lounkine}, \citenamefont {Keiser}, \citenamefont {Whitebread}, \citenamefont
  {Mikhailov}, \citenamefont {Hamon}, \citenamefont {Jenkins}, \citenamefont
  {Lavan}, \citenamefont {Weber}, \citenamefont {Doak}, \citenamefont
  {C{\^o}t{\'e}} \emph {et~al.}}]{lounkine2012large}%
  \BibitemOpen
  \bibfield  {author} {\bibinfo {author} {\bibfnamefont {E.}~\bibnamefont
  {Lounkine}}, \bibinfo {author} {\bibfnamefont {M.~J.}\ \bibnamefont
  {Keiser}}, \bibinfo {author} {\bibfnamefont {S.}~\bibnamefont {Whitebread}},
  \bibinfo {author} {\bibfnamefont {D.}~\bibnamefont {Mikhailov}}, \bibinfo
  {author} {\bibfnamefont {J.}~\bibnamefont {Hamon}}, \bibinfo {author}
  {\bibfnamefont {J.~L.}\ \bibnamefont {Jenkins}}, \bibinfo {author}
  {\bibfnamefont {P.}~\bibnamefont {Lavan}}, \bibinfo {author} {\bibfnamefont
  {E.}~\bibnamefont {Weber}}, \bibinfo {author} {\bibfnamefont {A.~K.}\
  \bibnamefont {Doak}}, \bibinfo {author} {\bibfnamefont {S.}~\bibnamefont
  {C{\^o}t{\'e}}},  \emph {et~al.},\ }\href@noop {} {\bibfield  {journal}
  {\bibinfo  {journal} {Nature}\ }\textbf {\bibinfo {volume} {486}},\ \bibinfo
  {pages} {361} (\bibinfo {year} {2012})}\BibitemShut {NoStop}%
\bibitem [{\citenamefont {Huigens~III}\ \emph {et~al.}(2013)\citenamefont
  {Huigens~III}, \citenamefont {Morrison}, \citenamefont {Hicklin},
  \citenamefont {Flood~Jr}, \citenamefont {Richter},\ and\ \citenamefont
  {Hergenrother}}]{huigens2013ring}%
  \BibitemOpen
  \bibfield  {author} {\bibinfo {author} {\bibfnamefont {R.~W.}\ \bibnamefont
  {Huigens~III}}, \bibinfo {author} {\bibfnamefont {K.~C.}\ \bibnamefont
  {Morrison}}, \bibinfo {author} {\bibfnamefont {R.~W.}\ \bibnamefont
  {Hicklin}}, \bibinfo {author} {\bibfnamefont {T.~A.}\ \bibnamefont
  {Flood~Jr}}, \bibinfo {author} {\bibfnamefont {M.~F.}\ \bibnamefont
  {Richter}}, \ and\ \bibinfo {author} {\bibfnamefont {P.~J.}\ \bibnamefont
  {Hergenrother}},\ }\href@noop {} {\bibfield  {journal} {\bibinfo  {journal}
  {Nature chemistry}\ }\textbf {\bibinfo {volume} {5}},\ \bibinfo {pages} {195}
  (\bibinfo {year} {2013})}\BibitemShut {NoStop}%
\bibitem [{\citenamefont {Todeschini}\ and\ \citenamefont
  {Consonni}(2009)}]{TodeschiniConsonniHandbookDescriptor}%
  \BibitemOpen
  \bibfield  {author} {\bibinfo {author} {\bibfnamefont {R.}~\bibnamefont
  {Todeschini}}\ and\ \bibinfo {author} {\bibfnamefont {V.}~\bibnamefont
  {Consonni}},\ }\href@noop {} {\emph {\bibinfo {title} {Handbook of Molecular
  Descriptors}}}\ (\bibinfo  {publisher} {Wiley-VCH, Weinheim},\ \bibinfo
  {year} {2009})\BibitemShut {NoStop}%
\bibitem [{\citenamefont {Faulon}\ \emph {et~al.}(2003)\citenamefont {Faulon},
  \citenamefont {{Visco, Jr.}},\ and\ \citenamefont
  {Pophale}}]{SignatureFaulon2003}%
  \BibitemOpen
  \bibfield  {author} {\bibinfo {author} {\bibfnamefont {J.-L.}\ \bibnamefont
  {Faulon}}, \bibinfo {author} {\bibfnamefont {D.~P.}\ \bibnamefont {{Visco,
  Jr.}}}, \ and\ \bibinfo {author} {\bibfnamefont {R.~S.}\ \bibnamefont
  {Pophale}},\ }\href@noop {} {\bibfield  {journal} {\bibinfo  {journal} {J.
  Chem. Inf. Comp. Sci.}\ }\textbf {\bibinfo {volume} {43}},\ \bibinfo {pages}
  {707} (\bibinfo {year} {2003})}\BibitemShut {NoStop}%
\bibitem [{\citenamefont {Visco}\ \emph {et~al.}(2002)\citenamefont {Visco},
  \citenamefont {Pophale}, \citenamefont {Rintoul},\ and\ \citenamefont
  {Faulon}}]{Visco2002}%
  \BibitemOpen
  \bibfield  {author} {\bibinfo {author} {\bibfnamefont {J.}~\bibnamefont
  {Visco}}, \bibinfo {author} {\bibfnamefont {R.~S.}\ \bibnamefont {Pophale}},
  \bibinfo {author} {\bibfnamefont {M.~D.}\ \bibnamefont {Rintoul}}, \ and\
  \bibinfo {author} {\bibfnamefont {J.~L.}\ \bibnamefont {Faulon}},\
  }\href@noop {} {\bibfield  {journal} {\bibinfo  {journal} {J. Mol. Graph.
  Model.}\ }\textbf {\bibinfo {volume} {20}},\ \bibinfo {pages} {429} (\bibinfo
  {year} {2002})}\BibitemShut {NoStop}%
\bibitem [{sup()}]{supplementary}%
  \BibitemOpen
  \href@noop {} {}\bibinfo {note} {See Supplemental Material}\BibitemShut
  {NoStop}%
\bibitem [{\citenamefont {O'Boyle}\ \emph {et~al.}(2011)\citenamefont
  {O'Boyle}, \citenamefont {Banck}, \citenamefont {James}, \citenamefont
  {Morley}, \citenamefont {Vandermeersch},\ and\ \citenamefont
  {Hutchison}}]{obabel}%
  \BibitemOpen
  \bibfield  {author} {\bibinfo {author} {\bibfnamefont {N.~M.}\ \bibnamefont
  {O'Boyle}}, \bibinfo {author} {\bibfnamefont {M.}~\bibnamefont {Banck}},
  \bibinfo {author} {\bibfnamefont {C.~A.}\ \bibnamefont {James}}, \bibinfo
  {author} {\bibfnamefont {C.}~\bibnamefont {Morley}}, \bibinfo {author}
  {\bibfnamefont {T.}~\bibnamefont {Vandermeersch}}, \ and\ \bibinfo {author}
  {\bibfnamefont {G.~R.}\ \bibnamefont {Hutchison}},\ }\href {\doibase
  10.1186/1758-2946-3-33} {\bibfield  {journal} {\bibinfo  {journal} {J.
  Cheminform.}\ }\textbf {\bibinfo {volume} {3}},\ \bibinfo {pages} {1}
  (\bibinfo {year} {2011})}\BibitemShut {NoStop}%
\bibitem [{\citenamefont {Landrum}(2014)}]{landrum2014rdkit}%
  \BibitemOpen
  \bibfield  {author} {\bibinfo {author} {\bibfnamefont {G.}~\bibnamefont
  {Landrum}},\ }\href@noop {} {\bibfield  {journal} {\bibinfo  {journal}
  {http://www.rdkit.org}\ }\textbf {\bibinfo {volume} {3}},\ \bibinfo {pages}
  {2012} (\bibinfo {year} {2014})}\BibitemShut {NoStop}%
\bibitem [{\citenamefont {Faber}\ \emph {et~al.}(2017)\citenamefont {Faber},
  \citenamefont {Hutchison}, \citenamefont {Huang}, \citenamefont {Gilmer},
  \citenamefont {Schoenholz}, \citenamefont {Dahl}, \citenamefont {Vinyals},
  \citenamefont {Kearnes}, \citenamefont {Riley},\ and\ \citenamefont {von
  Lilienfeld}}]{faber2017fast}%
  \BibitemOpen
  \bibfield  {author} {\bibinfo {author} {\bibfnamefont {F.~A.}\ \bibnamefont
  {Faber}}, \bibinfo {author} {\bibfnamefont {L.}~\bibnamefont {Hutchison}},
  \bibinfo {author} {\bibfnamefont {B.}~\bibnamefont {Huang}}, \bibinfo
  {author} {\bibfnamefont {J.}~\bibnamefont {Gilmer}}, \bibinfo {author}
  {\bibfnamefont {S.~S.}\ \bibnamefont {Schoenholz}}, \bibinfo {author}
  {\bibfnamefont {G.~E.}\ \bibnamefont {Dahl}}, \bibinfo {author}
  {\bibfnamefont {O.}~\bibnamefont {Vinyals}}, \bibinfo {author} {\bibfnamefont
  {S.}~\bibnamefont {Kearnes}}, \bibinfo {author} {\bibfnamefont {P.~F.}\
  \bibnamefont {Riley}}, \ and\ \bibinfo {author} {\bibfnamefont {O.~A.}\
  \bibnamefont {von Lilienfeld}},\ }\href@noop {} {\bibfield  {journal}
  {\bibinfo  {journal} {arXiv preprint arXiv:1702.05532}\ } (\bibinfo {year}
  {2017})}\BibitemShut {NoStop}%
\bibitem [{\citenamefont {M{\"u}ller}\ \emph {et~al.}(2001)\citenamefont
  {M{\"u}ller}, \citenamefont {Mika}, \citenamefont {R{\"a}tsch}, \citenamefont
  {Tsuda},\ and\ \citenamefont {Sch{\"o}lkopf}}]{muller2001introduction}%
  \BibitemOpen
  \bibfield  {author} {\bibinfo {author} {\bibfnamefont {K.-R.}\ \bibnamefont
  {M{\"u}ller}}, \bibinfo {author} {\bibfnamefont {S.}~\bibnamefont {Mika}},
  \bibinfo {author} {\bibfnamefont {G.}~\bibnamefont {R{\"a}tsch}}, \bibinfo
  {author} {\bibfnamefont {K.}~\bibnamefont {Tsuda}}, \ and\ \bibinfo {author}
  {\bibfnamefont {B.}~\bibnamefont {Sch{\"o}lkopf}},\ }\href@noop {} {\bibfield
   {journal} {\bibinfo  {journal} {IEEE transactions on neural networks}\
  }\textbf {\bibinfo {volume} {12}},\ \bibinfo {pages} {181} (\bibinfo {year}
  {2001})}\BibitemShut {NoStop}%
\bibitem [{\citenamefont {Sch{\"o}lkopf}\ and\ \citenamefont
  {Smola}(2002)}]{scholkopf2002learning}%
  \BibitemOpen
  \bibfield  {author} {\bibinfo {author} {\bibfnamefont {B.}~\bibnamefont
  {Sch{\"o}lkopf}}\ and\ \bibinfo {author} {\bibfnamefont {A.~J.}\ \bibnamefont
  {Smola}},\ }\href@noop {} {\emph {\bibinfo {title} {Learning with kernels:
  support vector machines, regularization, optimization, and beyond}}}\
  (\bibinfo  {publisher} {MIT press},\ \bibinfo {year} {2002})\BibitemShut
  {NoStop}%
\bibitem [{\citenamefont {Vovk}(2013)}]{Vovk2013}%
  \BibitemOpen
  \bibfield  {author} {\bibinfo {author} {\bibfnamefont {V.}~\bibnamefont
  {Vovk}},\ }\enquote {\bibinfo {title} {Kernel ridge regression},}\ in\ \href
  {\doibase 10.1007/978-3-642-41136-6_11} {\emph {\bibinfo {booktitle}
  {Empirical Inference: Festschrift in Honor of Vladimir N. Vapnik}}},\
  \bibinfo {editor} {edited by\ \bibinfo {editor} {\bibfnamefont
  {B.}~\bibnamefont {Sch{\"o}lkopf}}, \bibinfo {editor} {\bibfnamefont
  {Z.}~\bibnamefont {Luo}}, \ and\ \bibinfo {editor} {\bibfnamefont
  {V.}~\bibnamefont {Vovk}}}\ (\bibinfo  {publisher} {Springer Berlin
  Heidelberg},\ \bibinfo {address} {Berlin, Heidelberg},\ \bibinfo {year}
  {2013})\ pp.\ \bibinfo {pages} {105--116}\BibitemShut {NoStop}%
\bibitem [{\citenamefont {Hastie}\ \emph {et~al.}(2011)\citenamefont {Hastie},
  \citenamefont {Tibshirani},\ and\ \citenamefont
  {Friedman}}]{kernel_ridge_regression2}%
  \BibitemOpen
  \bibfield  {author} {\bibinfo {author} {\bibfnamefont {T.}~\bibnamefont
  {Hastie}}, \bibinfo {author} {\bibfnamefont {R.}~\bibnamefont {Tibshirani}},
  \ and\ \bibinfo {author} {\bibfnamefont {J.}~\bibnamefont {Friedman}},\
  }\href@noop {} {\emph {\bibinfo {title} {The Elements of Statistical
  Learning: Data Mining, Inference, and Prediction, Second Edition}}},\
  \bibinfo {edition} {2nd}\ ed.\ (\bibinfo  {publisher} {Springer},\ \bibinfo
  {address} {New York},\ \bibinfo {year} {2011})\BibitemShut {NoStop}%
\bibitem [{\citenamefont {Hoerl}\ \emph {et~al.}(2000)\citenamefont {Hoerl},
  \citenamefont {Arthur}, \citenamefont {Kennard},\ and\ \citenamefont
  {Robert}}]{Ridge_Regression}%
  \BibitemOpen
  \bibfield  {author} {\bibinfo {author} {\bibnamefont {Hoerl}}, \bibinfo
  {author} {\bibfnamefont {E.}~\bibnamefont {Arthur}}, \bibinfo {author}
  {\bibnamefont {Kennard}}, \ and\ \bibinfo {author} {\bibfnamefont
  {W.}~\bibnamefont {Robert}},\ }\href@noop {} {\bibfield  {journal} {\bibinfo
  {journal} {Technometrics}\ ,\ \bibinfo {pages} {80}} (\bibinfo {year}
  {2000})}\BibitemShut {NoStop}%
\bibitem [{\citenamefont {Pedregosa}\ \emph {et~al.}(2011)\citenamefont
  {Pedregosa}, \citenamefont {Varoquaux}, \citenamefont {Gramfort},
  \citenamefont {Michel}, \citenamefont {Thirion}, \citenamefont {Grisel},
  \citenamefont {Blondel}, \citenamefont {Prettenhofer}, \citenamefont {Weiss},
  \citenamefont {Dubourg}, \citenamefont {Vanderplas}, \citenamefont {Passos},
  \citenamefont {Cournapeau}, \citenamefont {Brucher}, \citenamefont {Perrot},\
  and\ \citenamefont {Duchesnay}}]{scikit-learn}%
  \BibitemOpen
  \bibfield  {author} {\bibinfo {author} {\bibfnamefont {F.}~\bibnamefont
  {Pedregosa}}, \bibinfo {author} {\bibfnamefont {G.}~\bibnamefont
  {Varoquaux}}, \bibinfo {author} {\bibfnamefont {A.}~\bibnamefont {Gramfort}},
  \bibinfo {author} {\bibfnamefont {V.}~\bibnamefont {Michel}}, \bibinfo
  {author} {\bibfnamefont {B.}~\bibnamefont {Thirion}}, \bibinfo {author}
  {\bibfnamefont {O.}~\bibnamefont {Grisel}}, \bibinfo {author} {\bibfnamefont
  {M.}~\bibnamefont {Blondel}}, \bibinfo {author} {\bibfnamefont
  {P.}~\bibnamefont {Prettenhofer}}, \bibinfo {author} {\bibfnamefont
  {R.}~\bibnamefont {Weiss}}, \bibinfo {author} {\bibfnamefont
  {V.}~\bibnamefont {Dubourg}}, \bibinfo {author} {\bibfnamefont
  {J.}~\bibnamefont {Vanderplas}}, \bibinfo {author} {\bibfnamefont
  {A.}~\bibnamefont {Passos}}, \bibinfo {author} {\bibfnamefont
  {D.}~\bibnamefont {Cournapeau}}, \bibinfo {author} {\bibfnamefont
  {M.}~\bibnamefont {Brucher}}, \bibinfo {author} {\bibfnamefont
  {M.}~\bibnamefont {Perrot}}, \ and\ \bibinfo {author} {\bibfnamefont
  {E.}~\bibnamefont {Duchesnay}},\ }\href@noop {} {\bibfield  {journal}
  {\bibinfo  {journal} {J. Mach. Learn. Res.}\ }\textbf {\bibinfo {volume}
  {12}},\ \bibinfo {pages} {2825} (\bibinfo {year} {2011})}\BibitemShut
  {NoStop}%
\bibitem [{\citenamefont {Neal}(1996)}]{Neal:1996:BLN:525544}%
  \BibitemOpen
  \bibfield  {author} {\bibinfo {author} {\bibfnamefont {R.~M.}\ \bibnamefont
  {Neal}},\ }\href@noop {} {\emph {\bibinfo {title} {Bayesian Learning for
  Neural Networks}}}\ (\bibinfo  {publisher} {Springer-Verlag New York, Inc.},\
  \bibinfo {address} {Secaucus, NJ, USA},\ \bibinfo {year} {1996})\BibitemShut
  {NoStop}%
\bibitem [{\citenamefont {Zou}\ and\ \citenamefont
  {Hastie}(2005)}]{RSSB:RSSB503}%
  \BibitemOpen
  \bibfield  {author} {\bibinfo {author} {\bibfnamefont {H.}~\bibnamefont
  {Zou}}\ and\ \bibinfo {author} {\bibfnamefont {T.}~\bibnamefont {Hastie}},\
  }\href@noop {} {\bibfield  {journal} {\bibinfo  {journal} {J. R. Stat. Soc.
  Series. B Stat. Methodol.}\ }\textbf {\bibinfo {volume} {67}},\ \bibinfo
  {pages} {301} (\bibinfo {year} {2005})}\BibitemShut {NoStop}%
\bibitem [{\citenamefont {Breiman}(2001)}]{breiman2001random}%
  \BibitemOpen
  \bibfield  {author} {\bibinfo {author} {\bibfnamefont {L.}~\bibnamefont
  {Breiman}},\ }\href@noop {} {\bibfield  {journal} {\bibinfo  {journal}
  {Machine learning}\ }\textbf {\bibinfo {volume} {45}},\ \bibinfo {pages} {5}
  (\bibinfo {year} {2001})}\BibitemShut {NoStop}%
\bibitem [{\citenamefont {Duvenaud}\ \emph {et~al.}(2015)\citenamefont
  {Duvenaud}, \citenamefont {Maclaurin}, \citenamefont {Iparraguirre},
  \citenamefont {Bombarell}, \citenamefont {Hirzel}, \citenamefont
  {Aspuru-Guzik},\ and\ \citenamefont {Adams}}]{duvenaud2015convolutional}%
  \BibitemOpen
  \bibfield  {author} {\bibinfo {author} {\bibfnamefont {D.~K.}\ \bibnamefont
  {Duvenaud}}, \bibinfo {author} {\bibfnamefont {D.}~\bibnamefont {Maclaurin}},
  \bibinfo {author} {\bibfnamefont {J.}~\bibnamefont {Iparraguirre}}, \bibinfo
  {author} {\bibfnamefont {R.}~\bibnamefont {Bombarell}}, \bibinfo {author}
  {\bibfnamefont {T.}~\bibnamefont {Hirzel}}, \bibinfo {author} {\bibfnamefont
  {A.}~\bibnamefont {Aspuru-Guzik}}, \ and\ \bibinfo {author} {\bibfnamefont
  {R.~P.}\ \bibnamefont {Adams}},\ }in\ \href@noop {} {\emph {\bibinfo
  {booktitle} {Advances in Neural Information Processing Systems}}}\ (\bibinfo
  {year} {2015})\ pp.\ \bibinfo {pages} {2215--2223}\BibitemShut {NoStop}%
\bibitem [{\citenamefont {Desautels}\ \emph {et~al.}(2014)\citenamefont
  {Desautels}, \citenamefont {Krause},\ and\ \citenamefont
  {Burdick}}]{JMLR:v15:desautels14a}%
  \BibitemOpen
  \bibfield  {author} {\bibinfo {author} {\bibfnamefont {T.}~\bibnamefont
  {Desautels}}, \bibinfo {author} {\bibfnamefont {A.}~\bibnamefont {Krause}}, \
  and\ \bibinfo {author} {\bibfnamefont {J.~W.}\ \bibnamefont {Burdick}},\
  }\href@noop {} {\bibfield  {journal} {\bibinfo  {journal} {J. Mach. Learn.
  Res.}\ }\textbf {\bibinfo {volume} {15}},\ \bibinfo {pages} {4053} (\bibinfo
  {year} {2014})}\BibitemShut {NoStop}%
\bibitem [{hyp(2016)}]{hypertune}%
  \BibitemOpen
  \href@noop {} {\enquote {\bibinfo {title} {Google hypertune},}\ }\bibinfo
  {howpublished} {https://cloud.google.com/ml/} (\bibinfo {year} {Accessed:
  2016})\BibitemShut {NoStop}%
\bibitem [{\citenamefont {Tew}\ \emph {et~al.}(2007)\citenamefont {Tew},
  \citenamefont {Klopper}, \citenamefont {Heckert},\ and\ \citenamefont
  {Gauss}}]{Wim2007Harmonic}%
  \BibitemOpen
  \bibfield  {author} {\bibinfo {author} {\bibfnamefont {D.~P.}\ \bibnamefont
  {Tew}}, \bibinfo {author} {\bibfnamefont {W.}~\bibnamefont {Klopper}},
  \bibinfo {author} {\bibfnamefont {M.}~\bibnamefont {Heckert}}, \ and\
  \bibinfo {author} {\bibfnamefont {J.}~\bibnamefont {Gauss}},\ }\href
  {\doibase 10.1021/jp070851u} {\bibfield  {journal} {\bibinfo  {journal} {J.
  Phys. Chem. A}\ }\textbf {\bibinfo {volume} {111}},\ \bibinfo {pages} {11242}
  (\bibinfo {year} {2007})}\BibitemShut {NoStop}%
\bibitem [{\citenamefont {M{\"u}ller}\ \emph {et~al.}(1996)\citenamefont
  {M{\"u}ller}, \citenamefont {Finke}, \citenamefont {Murata}, \citenamefont
  {Schulten},\ and\ \citenamefont {Amari}}]{Muller1996}%
  \BibitemOpen
  \bibfield  {author} {\bibinfo {author} {\bibfnamefont {K.-R.}\ \bibnamefont
  {M{\"u}ller}}, \bibinfo {author} {\bibfnamefont {M.}~\bibnamefont {Finke}},
  \bibinfo {author} {\bibfnamefont {N.}~\bibnamefont {Murata}}, \bibinfo
  {author} {\bibfnamefont {K.}~\bibnamefont {Schulten}}, \ and\ \bibinfo
  {author} {\bibfnamefont {S.}~\bibnamefont {Amari}},\ }\href@noop {}
  {\bibfield  {journal} {\bibinfo  {journal} {Neural Comput.}\ }\textbf
  {\bibinfo {volume} {8}},\ \bibinfo {pages} {1085} (\bibinfo {year}
  {1996})}\BibitemShut {NoStop}%
\bibitem [{\citenamefont {Gilmer}\ \emph {et~al.}(2017)\citenamefont {Gilmer},
  \citenamefont {Schoenholz}, \citenamefont {Riley}, \citenamefont {Vinyals},\
  and\ \citenamefont {Dahl}}]{NeuralMessagePassing}%
  \BibitemOpen
  \bibfield  {author} {\bibinfo {author} {\bibfnamefont {J.}~\bibnamefont
  {Gilmer}}, \bibinfo {author} {\bibfnamefont {S.~S.}\ \bibnamefont
  {Schoenholz}}, \bibinfo {author} {\bibfnamefont {P.~F.}\ \bibnamefont
  {Riley}}, \bibinfo {author} {\bibfnamefont {O.}~\bibnamefont {Vinyals}}, \
  and\ \bibinfo {author} {\bibfnamefont {G.~E.}\ \bibnamefont {Dahl}},\ }in\
  \href@noop {} {\emph {\bibinfo {booktitle} {Proceedings of the 34nd
  International Conference on Machine Learning, {ICML} 2017}}}\ (\bibinfo
  {year} {2017})\BibitemShut {NoStop}%
\bibitem [{\citenamefont {Kirkpatrick}\ and\ \citenamefont
  {Ellis}(2004)}]{ChemicalSpace}%
  \BibitemOpen
  \bibfield  {author} {\bibinfo {author} {\bibfnamefont {P.}~\bibnamefont
  {Kirkpatrick}}\ and\ \bibinfo {author} {\bibfnamefont {C.}~\bibnamefont
  {Ellis}},\ }\href@noop {} {\bibfield  {journal} {\bibinfo  {journal}
  {Nature}\ }\textbf {\bibinfo {volume} {432}},\ \bibinfo {pages} {823}
  (\bibinfo {year} {2004})}\BibitemShut {NoStop}%
\bibitem [{\citenamefont {{G}hasemi}\ \emph {et~al.}(2010)\citenamefont
  {{G}hasemi}, \citenamefont {{A}msler}, \citenamefont {{H}ennig},
  \citenamefont {{R}oy}, \citenamefont {{G}oedecker}, \citenamefont
  {{L}enosky}, \citenamefont {{U}mrigar}, \citenamefont {{G}enovese},
  \citenamefont {{M}orishita},\ and\ \citenamefont {{N}ishio}}]{Goedecker2010}%
  \BibitemOpen
  \bibfield  {author} {\bibinfo {author} {\bibfnamefont {S.~A.}\ \bibnamefont
  {{G}hasemi}}, \bibinfo {author} {\bibfnamefont {M.}~\bibnamefont {{A}msler}},
  \bibinfo {author} {\bibfnamefont {R.~G.}\ \bibnamefont {{H}ennig}}, \bibinfo
  {author} {\bibfnamefont {S.}~\bibnamefont {{R}oy}}, \bibinfo {author}
  {\bibfnamefont {S.}~\bibnamefont {{G}oedecker}}, \bibinfo {author}
  {\bibfnamefont {T.~J.}\ \bibnamefont {{L}enosky}}, \bibinfo {author}
  {\bibfnamefont {C.~J.}\ \bibnamefont {{U}mrigar}}, \bibinfo {author}
  {\bibfnamefont {L.}~\bibnamefont {{G}enovese}}, \bibinfo {author}
  {\bibfnamefont {T.}~\bibnamefont {{M}orishita}}, \ and\ \bibinfo {author}
  {\bibfnamefont {K.}~\bibnamefont {{N}ishio}},\ }\href {\doibase
  10.1103/PhysRevB.81.214107} {\bibfield  {journal} {\bibinfo  {journal} {Phys.
  Rev. B}\ }\textbf {\bibinfo {volume} {81}},\ \bibinfo {pages} {214107}
  (\bibinfo {year} {2010})}\BibitemShut {NoStop}%
\end{thebibliography}%
\end{document}